\documentclass[preprint]{emulateapj}

\def\la{\lambda}

\usepackage[dvips]{color}

\usepackage{aas_macros} 
\usepackage{here}
\usepackage{amsmath,amssymb}
\slugcomment{}
\shortauthors{Hirano et al.}
\shorttitle{Measurements of Stellar Inclinations for KOI Systems II}
\begin{document}
\title{Measurements of Stellar Inclinations for Kepler Planet Candidates II:
Candidate Spin-Orbit Misalignments in Single and Multiple-Transiting Systems}
\author{
Teruyuki Hirano\altaffilmark{1}, 
Roberto Sanchis-Ojeda\altaffilmark{2},
Yoichi Takeda\altaffilmark{3},
Joshua N.\ Winn\altaffilmark{2},
Norio Narita\altaffilmark{3}, 
and
Yasuhiro H. Takahashi\altaffilmark{3,4}
} 
\altaffiltext{1}{Department of Earth and Planetary Sciences, Tokyo Institute of Technology,
2-12-1 Ookayama, Meguro-ku, Tokyo 152-8551, Japan
}
\altaffiltext{2}{Department of Physics, and Kavli Institute 
for Astrophysics and Space Research, Massachusetts Institute of Technology,
Cambridge, MA 02139}
\altaffiltext{3}{National Astronomical Observatory of Japan, 
2-21-1 Osawa, Mitaka, Tokyo, 181-8588, Japan}
\altaffiltext{4}{Department of Astronomy, The University of Tokyo, 
Tokyo, 113-0033, Japan}

\email{hirano@geo.titech.ac.jp}
\begin{abstract}

  We present a test for spin-orbit alignment for the host stars of 25
  candidate planetary systems detected by the {\it Kepler} spacecraft. The
  inclination angle of each star's rotation axis was estimated from
  its rotation period, rotational line broadening, and radius.  The
  rotation periods were determined using the {\it Kepler} photometric
  time series. The rotational line broadening was determined from
  high-resolution optical spectra with Subaru/HDS.  Those same spectra
  were used to determine the star's photospheric parameters (effective
  temperature, surface gravity, metallicity) which were then
  interpreted with stellar-evolutionary models to determine stellar
  radii.  We combine the new sample with the 7 stars from our previous
  work on this subject, finding that the stars show a statistical
  tendency to have inclinations near 90$^\circ$, in alignment with the
  planetary orbits.  Possible spin-orbit misalignments 
  are seen in several systems,
  including three multiple-planet systems (KOI-304, 988, 2261).
  Ideally these systems should be scrutinized with complementary
  techniques---such as the Rossiter-McLaughlin effect,
  starspot-crossing anomalies or asteroseismology---but the
  measurements will be difficult owing to the relatively faint
  apparent magnitudes and small transit signals in these systems.

\end{abstract}

\section{Introduction\label{s:intro}}\label{s:intro}

The angle of the stellar spin axis with respect to the planetary
orbital axis (spin-orbit angle) is an observable quantity that may be
important for understanding the evolutionary history of exoplanetary
systems. In order to explain the existence of close-in giant planets
(hot Jupiters or Neptunes), various migration scenarios have been
proposed, which differ in their predictions for the spin-orbit angle.
Some theories, such as disk migration, predict that the stellar spin
and planetary orbital axes should be well aligned
\citep[e.g.,][]{1996Natur.380..606L}. Other theories, such as planet-planet
scattering or Kozai migration, predict a very wide range of spin-orbit
angles \citep[see, e.g.,][]{2003ApJ...589..605W, 2011ApJ...742...72N,
  2007ApJ...669.1298F}.

Most of the current measurements of the spin-orbit angle have been
based on observations of the Rossiter-McLaughlin (RM) effect
\citep[e.g.,][]{2000A&A...359L..13Q, 2005ApJ...622.1118O,
  2005ApJ...631.1215W, 2007PASJ...59..763N, 2007ApJ...667..549W,
  2011ApJ...742...69H} or photometric anomalies due to transits over
starspots \citep[e.g.,][]{2011ApJ...733..127S, 2011ApJ...740L..10N, 2011ApJS..197...14D}.  
These measurements have revealed a diversity of spin-orbit angles
\citep[e.g.,][]{2008A&A...488..763H, 2009ApJ...703L..99W,
  2009PASJ...61L..35N}.  This diversity has inspired many theoretical
studies of the possible reasons for highly inclined planetary orbits
\citep[e.g.,][]{2011MNRAS.412.2790L, 2011Natur.473..187N}.  The
measurements have also revealed some possible patterns relating the
spin-orbit angle and the properties of the host stars
\citep{2010ApJ...718L.145W, 2012ApJ...757...18A}.  However, the
existing measurements have been almost exclusively restricted to
close-in giant planets. This is simply because the preceding
measurement techniques are best suited to relatively large planets,
which produce stronger spectroscopic or photometric signals during a
planetary transit. Thus, the spin-orbit relations for smaller planets
have been unknown until recently.

An important step was taken by \citet{2010ApJ...719..602S}, who
demonstrated that the stellar inclination angle (the angle between the
stellar spin axis and the line of sight) can be readily estimated for
a large number of transiting exoplanetary systems, and used to probe
spin-orbit alignment. The basic idea is to use estimates of the
rotation velocity $V$ and the projected rotation velocity $V\sin I_s$
to determine $\sin I_s$. Since the orbital axis of a transiting planet
must be nearly perpendicular to the line-of-sight ($\sin I_o \approx
1$), a small value of $\sin I_s$ implies a spin-orbit misalignment.

The pioneering analysis of \citet{2010ApJ...719..602S} was based on
spectroscopic determinations of $V\sin I_s$, as well as statistical estimates of $V$
based on the rotation-age-mass correlations that are observed for
main-sequence stars.  It is also possible to measure $V$ more
directly, if accurate estimates of the stellar radius $R_s$ and
rotation period $P_s$ are available, using the relation $V = 2\pi
R_s/P_s$ \citep[see, e.g.,][]{1984ApJ...287..307D}. 
It has also become possible to estimate $\sin I_s$ using
asteroseismology \citep[e.g.,][]{2013ApJ...766..101C}.

An important advantage of this technique is that the difficulty of
measuring stellar inclinations is independent of the size of the
transiting planet, and therefore the spin-orbit relation may be
investigated even for smaller planets (such as Earth-sized planets).
One shortcoming of this technique is that the relative uncertainty in
$I_s$ becomes large when $I_s$ approaches $90^\circ$. Another is that
it is often difficult to obtain accurate and precise measurements of
$V\sin I_s$ for cool stars ($T_{\rm eff}<6000$~K), for which the
rotational line broadening is often comparable to the effects of
instrumental broadening and macroturbulence.  This is contrast with
measurements of the RM effect, by which the sky-projected spin-orbit
angle $\la$ can often be measured to within 5-10 degrees
\citep[e.g.,][]{2010A&A...524A..25T, 2011PASJ...63L..57H}. For these
reasons, it may be best to regard this technique as an efficient
method for identifying low-inclination hot stars; and for identifying
{\it candidate} low-inclination cool stars that can be followed up
with complementary techniques.

In the precursor to this paper, \citet{2012ApJ...756...66H} determined
stellar inclinations for 7 host stars of transiting-planet candidates.
To measure rotation periods $P_s$, they used a periodogram analysis of
the light curve modulations seen with the {\it Kepler} telescope.
They also undertook new spectroscopic measurements of $V\sin I_s$ and
stellar radii $R_s$ via $I_s=\arcsin(P_s\cdot V\sin I_s/2\pi R_s)$,
for several KOIs (Kepler Objects of Interest).  They found that most
of the systems are consistent with $I_s=90^\circ$, suggesting good
spin-orbit alignment, but at least one system (KOI-261) may have a
spin-orbit misalignment. 
The planet Kepler-63b was also found to have a tilted orbit 
using the same technique, and also through the measurement of the sky-projected 
obliquity using the RM effect \citep{2013ApJ...775...54S}.
More recently, \citet{2013MNRAS.tmp.2426W}
applied the same technique and found candidate spin-orbit
misalignments for several KOI's including a multiple transiting system
(Kepler-9). Even more recently, a robust spin-orbit misalignment
around multiple systems was reported for Kepler-56 based on the
asteroseismic determination of the stellar inclination
\citep{2013arXiv1310.4503H}.

In this paper, we continue the effort by \citet{2012ApJ...756...66H}
to examine the stellar inclinations for KOI systems.  In the next
section, we describe the new spectroscopic observations with the
Subaru telescope to obtain basic spectroscopic parameters for 25 KOI
systems, including 10 systems with multiple transiting planets.  We
then present the analyses of stellar rotational periods and
spectroscopic parameters such as $V\sin I_s$ and $R_s$ in Section
\ref{s:result}.  Section \ref{s:discussion} presents a statistical
analysis of the observed distribution of $I_s$. We try to test some
hypotheses such as whether the observed values of $I_s$ are drawn from
an isotropic distribution (\S \ref{s:test}). Section \ref{s:summary}
summarizes our results and their implications.


\section{Target Selection and Observations \label{s:obs}}\label{s:obs}

We composed a list of KOIs for measurements of stellar inclinations
based on the following criteria: (1) a preliminary light curve
analysis shows a peak power in the Lomb-Scargle periodogram larger
than 1000, (2) the estimated rotation velocity at the stellar equator
is larger than about 3~km~s$^{-1}$, and (3) the apparent magnitude in
the \textit{Kepler} bandpass is $m_{\rm Kep} \lesssim 14$. The rotational
velocity needed for the second criterion was estimated from the
stellar radius in the Kepler Input Catalog (KIC) and the preliminary
estimate of the rotation period. We excluded slow rotators because the
measurement of $V\sin I_s$ for slow rotators ($V\sin I_s \lesssim
3$~km~s$^{-1}$) has a large fractional uncertainty, as shown below.

\begin{figure}[t]
\begin{center}
\includegraphics[width=9cm,clip]{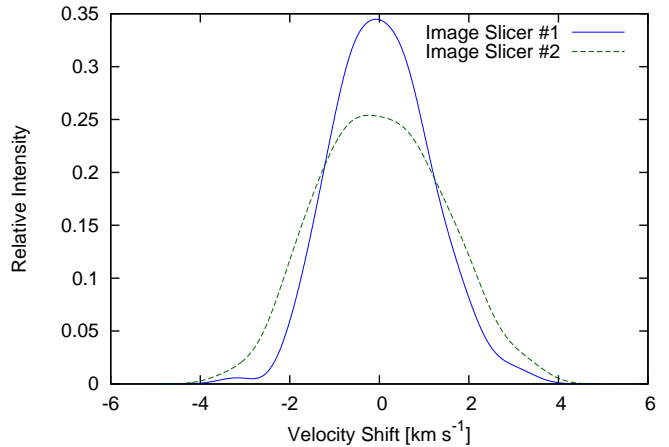} 
\caption{Instrumental profiles of Subaru/HDS for Image Slicer \#1 (blue solid line) and 
\#2 (green dashed line). These profiles were extracted from the same
spectral region that
was used to determine the $V\sin I_s$ of the program stars.
}\label{fig:IP}
\end{center}
\end{figure}

In order to estimate the basic spectroscopic parameters, we conducted
high dispersion spectroscopy with Subaru/HDS on 2012 June 30, July 1,
2, and September 4; and on 2013~June 20 and 21. All together we
obtained spectra for 25 KOIs. During the 2012 observations, we
employed the standard ``I2a" setup with Image Slicer \#1
\citep{2012PASJ...64...77T}, attaining a spectral resolution of $R\sim
110,000$.  For the 2013 observations we used Image Slicer \#2 ($R\sim
80,000$). On each night of observations, we obtained a spectrum of the
flat-field lamp through the iodine cell, to determine the instrumental
line broadening function.  In the subsequent analysis, the line
broadening due to the instrumental profile (IP) for each setup was
deconvolved as shown in Figure \ref{fig:IP}, and taken into account
when we estimated the rotation velocity of each star.

Each spectrum was subjected to standard IRAF procedures to extract a
one-dimensional (1D) spectrum. The wavelength scale was set with
reference to a spectrum of the thorium-argon lamp. The resultant
signal-to-noise ratio (SNR) in the 1D spectrum was typically
50-100~pixel$^{-1}$.  The I2a setup covers the spectral region between
4900-7600~$\mathrm{\AA}$, within which there is a large number of iron
lines available for estimation of the photospheric parameters.

\section{Analyses and Results \label{s:result}}\label{s:result}
\subsection{Estimate for Rotation Periods}\label{s:periods_estimate}

We determined the rotation periods of the stars using the photometric
observations provided by the \textit{Kepler} telescope
\citep{2010Sci...327..977B}. In particular, we used the Long Cadence
data (30~minute integrations) available from the MAST archive from
quarters 2 through 16, for up to a total of approximately 4 years of
data. Previously, \citet{2012ApJ...756...66H} used the simple aperture
flux data to obtain the rotation periods, but those data needed to be
treated carefully to remove systematic and instrumental effects on
timescales similar to the rotation periods. In this paper we used the
PDC-MAP final data product, since it is designed to remove the
unphysical trends leaving the signal of stellar spots unaltered
\citep{2012PASP..124.1000S, 2012PASP..124..985S}.
 
To excise the data obtained during transits, we identified the transit
intervals using the publicly available transit ephemerides
\citep{2013ApJS..204...24B} downloaded from the NASA exoplanet archive
\citep{2013PASP..125..989A}, which are based on the assumption of
constant orbital periods. We also removed gross outliers, and
normalized the data from each quarter by dividing by the quarterly
median flux.  We then computed the Lomb-Scargle periodogram 
adopting the definition and algorithm described by \citet{1989ApJ...338..277P}.
In general each periodogram showed several peaks, the strongest of which
can be attributed to stellar variability.  We selected the strongest
peak of the periodogram as the first candidate for the rotation period, and
adopted the full width at half maximum (FWHM) of the peak as the
1$\sigma$ uncertainty. We also performed a visual inspection of each light 
curve to make sure that the stellar flux appeared to be varying quasi-periodically with
the candidate rotation period, as opposed to a more regular periodic
signal that would be caused by orbital effects or pulsation. In
particular, we looked for quasi-sinusoidal variations with slow
amplitude and phase modulation on a timescale of a few rotation
periods, as would be expected of starspots.  We also checked that
there was not additional power at twice the candidate rotation period, as it
sometimes happens when a star has two similar size starspots in
opposite longitudes.  Such a configuration causes the flux variations
to peak twice per rotation period, inducing a substitute for a
subharmonic peak at
half the rotation period, which in some occasions could be more
significant than the real rotation period peak, making our code
identify the wrong rotation period. In two cases, KOI-180 and KOI-2636, 
the strongest peak corresponded to half the rotation period, 
so we matched the correct peak with the rotation period, and 
assigned the right uncertainty neglecting all the power at half the rotation period. 
Table~\ref{tab:rot} summarizes our rotation
period measurements, including the peak value of the periodogram power
and the variability amplitude, defined as the full range of flux after
eliminating the lowest 10\% and the highest 10\% of the flux values
\citep{2012ApJ...756...66H}.

\begin{figure}[t]
\begin{center}
\includegraphics[width=9cm,clip]{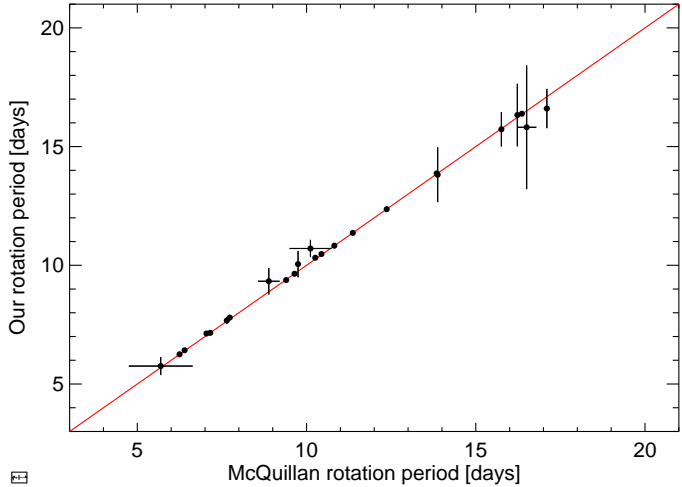} 
\caption{Comparison between the rotation periods listed in Table \ref{tab:rot} and
periods estimated by \citet{2013ApJ...775L..11M}. 
}\label{fig:checkperiod}
\end{center}
\end{figure}
\citet{2013MNRAS.432.1203M} advocated the autocorrelation function,
rather than the Lomb-Scargle periodogram, for measuring rotation
periods with \textit{Kepler} data.  We checked our measured rotation periods
against a published table of rotation periods that were determined
using the autocorrelation function \citep{2013ApJ...775L..11M}, and
found good agreement between the results of both techniques (Figure
\ref{fig:checkperiod}), although our quoted uncertainties are always
larger.

\begin{table}[t]
\caption{
Rotation periods estimated from the \textit{Kepler} photometry. 
Also given are the peak periodogram power and the
variability amplitude (defined as in the text).
}\label{tab:rot}
\begin{tabular}{cccc}
\hline
System & $P_s$ (days)& Peak Power & Variability Amplitude (\%)\\\hline\hline
KOI-180 & $15.728 \pm 0.726$ & 3266.21 & 0.348\\ 
KOI-285 & $16.829 \pm 0.588$ & 1474.14 & 0.013\\ 
KOI-304 & $15.814 \pm 2.606$ & 1470.75 & 0.070\\ 
KOI-323 & $7.674 \pm 0.143$ & 4121.72 & 0.566\\ 
KOI-635 & $9.328 \pm 0.558$ & 1814.23 & 0.142\\ 
KOI-678 & $13.871 \pm 0.058$ & 14941.34 & 0.924\\ 
KOI-718 & $16.603 \pm 0.828$ & 1225.87 & 0.045\\ 
KOI-720 & $9.378 \pm 0.027$ & 7967.45 & 0.670\\ 
KOI-988 & $12.363 \pm 0.064$ & 15540.17 & 0.685\\ 
KOI-1615 & $7.797 \pm 0.043$ & 5516.86 & 0.256\\ 
KOI-1628 & $5.756 \pm 0.378$ & 2540.79 & 0.179\\ 
KOI-1779 & $7.154 \pm 0.014$ & 9393.85 & 0.548\\ 
KOI-1781 & $10.474 \pm 0.084$ & 7000.01 & 0.733\\ 
KOI-1797 & $10.826 \pm 0.033$ & 14184.33 & 0.679\\ 
KOI-1835 & $9.644 \pm 0.028$ & 11825.34 & 0.531\\ 
KOI-1839 & $6.252 \pm 0.027$ & 8156.72 & 0.810\\ 
KOI-1890 & $6.420 \pm 0.039$ & 2716.78 & 0.020\\ 
KOI-1916 & $10.318 \pm 0.097$ & 3128.06 & 0.144\\ 
KOI-2001 & $16.385 \pm 0.081$ & 14017.50 & 0.843\\ 
KOI-2002 & $10.708 \pm 0.364$ & 1989.29 & 0.138\\ 
KOI-2026 & $10.051 \pm 0.555$ & 2530.09 & 0.191\\ 
KOI-2035 & $7.127 \pm 0.104$ & 7347.71 & 0.741\\ 
KOI-2087 & $13.816 \pm 1.155$ & 1902.26 & 0.086\\ 
KOI-2261 & $11.366 \pm 0.042$ & 17009.04 & 0.515\\ 
KOI-2636 & $16.330 \pm 1.317$ & 1774.31 & 0.077\\
\hline
\end{tabular}
\end{table}

\subsection{Spectroscopic Parameters}

\subsubsection{Photospheric Parameters, and Stellar Radius\label{sec:atmospheric}}\label{sec:atmospheric}

Based on \citet{2002PASJ...54..451T, 2005PASJ...57...27T}, we
estimated the basic photospheric parameters (the effective temperature
$T_\mathrm{\rm eff}$, surface gravity $\log g$, microturbulent
velocity $\xi$, and metallicity [Fe/H]) by measuring the equivalent
widths of the available iron absorption lines.
That is, these parameters are established by requiring that
the following three conditions are simultaneously fulfilled:
(a) excitation equilibrium (Fe abundances show no systematic
dependence on the excitation potential), (b) ionization equilibrium
(mean Fe abundance from Fe I lines and that from Fe II lines
agree with each other), and (c) curve-of-growth matching (Fe
abundances do not systematically depend on line strengths).
We used typically $\sim$150--200 and $\sim$10--15 lines
for Fe I and Fe II, respectively.

We next convert the photospheric parameters into stellar masses and radii 
employing the Yonsei-Yale (Y$^2$) stellar-evolutionary models \citep{2001ApJS..136..417Y}.  
Since an accurate estimation of the stellar radius is essential in our methodology,
it is important to take account of the accuracy of the photospheric parameters.  
\citet{2010MNRAS.405.1907B} spectroscopically analyzed 23 solar-type stars, 
arguing that the ``true" effective temperature 
of a star defined from the stellar luminosity and radius might have a systematic offset of
$-40\pm 20$ K  from the spectroscopic model parameter $T_\mathrm{eff}$,
while spectroscopic measurements of the surface gravity $\log g$ did not show 
a significant offset. 
Since our spectroscopic measurement of $T_\mathrm{eff}$ is similar to that of
\citet{2010MNRAS.405.1907B}, we assume that the systematic error in $T_\mathrm{eff}$
is 40 K, which is quadratically added to the internal statistical error listed in Table \ref{table2}
when we estimate the stellar radii and masses based with the Y$^2$ isochrones.
To account for these uncertainties (both statistic and systematic) in the photospheric  parameters, 
we randomly generated many sets of ($T_\mathrm{\rm eff},~\log g$, and [Fe/H]) 
assuming Gaussian distributions for their uncertainties. 
Each set of ($T_\mathrm{\rm eff}, ~\log g$, and [Fe/H]) was then converted 
to the mass and radius on the Y$^2$ isochrones.
The resultant distributions give the estimates (and errors) for the mass and radius of each system.  
Table \ref{table2} summarizes our measurements of the photospheric parameters together
with the stellar radius.

\begin{figure}[t]
\begin{center}
\includegraphics[width=9cm,clip]{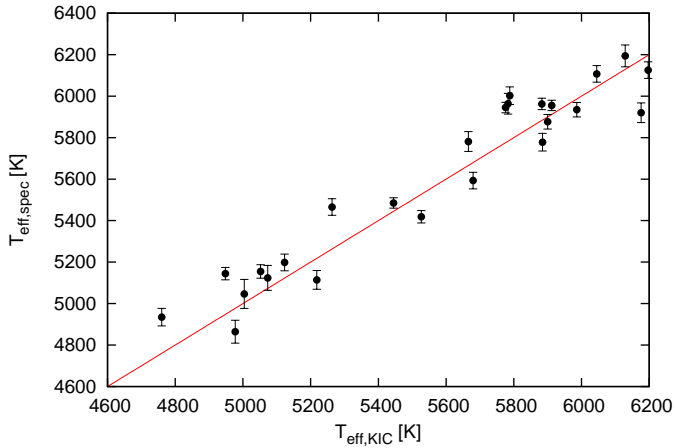} 
\caption{Comparison between our measurement of $T_\mathrm{\rm eff}$, and
the value of $T_\mathrm{\rm eff}$ reported in the Kepler Input Catalog (KIC). 
}\label{fig:Teff}
\end{center}
\end{figure}
\begin{figure}[t]
\begin{center}
\includegraphics[width=9cm,clip]{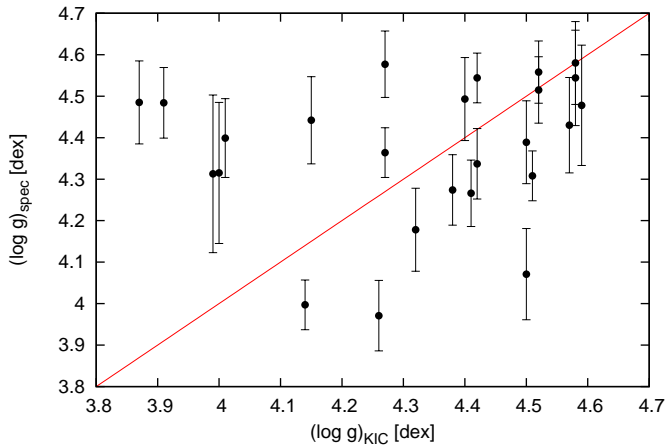} 
\caption{Comparison between our measurement of $\log g$ and
the KIC value of $\log g$. While both quantities are in good agreement for 
$\log g\gtrsim 4.3$~dex, large discrepancies are seen for smaller
value of $\log g$. 
}\label{fig:logg}
\end{center}
\end{figure}
To check whether our spectroscopically-derived photospheric parameters
are compatible with the parameters that were determined from broadband
photometry, we compared our effective temperatures and surface
gravities with the values reported in the Kepler Input Catalog (KIC).
Figures~\ref{fig:Teff} and~\ref{fig:logg} show these comparisons.  The
root-mean-squared residual between the spectroscopic and photometric
$T_\mathrm{\rm eff}$ and $\log g$ are $124$~K and $0.26$~dex,
respectively.  This level of agreement seems reasonable given the
relatively large uncertainties in the KIC parameters \citep[$\sim 200$
K for $T_\mathrm{eff}$ and $\sim 0.4$ dex for $\log
g$,][]{2011AJ....142..112B}.

\subsubsection{Projected Rotational Velocity\label{vsini_measure}}

We measured the projected rotational velocity $V\sin I_s$ by fitting a
model to the observed spectrum for each system. Theoretically, an
observed stellar spectrum $I_\mathrm{obs}(\la)$ can be considered as the
convolution of several functions:
\begin{eqnarray}
I_\mathrm{obs}(\la)=S(\la)*M(\la)*\mathrm{IP},
\label{model}
\end{eqnarray}
where $S(\la)$ is the intrinsic stellar spectrum taking into account
only thermal and natural broadening (including microturbulence),
$M(\la)$ is the broadening kernel representing rotation and
macroturbulence \citep{2005oasp.book.....G}, and IP represents the
instrumental line profile (see Figure \ref{fig:IP}). The IP was
determined by deconvolving the spectrum of the flat-field lamp through
the iodine cell. 
For each target star, we generated the intrinsic
spectrum $S(\la)$ based on the ATLAS9 model \citep[a plane-parallel stellar atmosphere 
model in LTE,][]{1993KurCD..13.....K}
with the input photospheric parameters being the best-fit values derived above,
and fitted the observed spectrum $I_\mathrm{obs}(\la)$, allowing $V\sin I_s$
to be a free parameter (which affects $M(\la)$).  As for the
macroturbulence, we adopted the radial-tangential model of
\citet{2005oasp.book.....G} and assumed that the macroturbulent
velocity $\zeta_\mathrm{RT}$ is expressed by the following empirical
formula \citep{2005ApJS..159..141V}:
\begin{eqnarray}
\zeta_\mathrm{RT}
=\left(3.98+\frac{T_\mathrm{eff}-5770~\mathrm{K}}{650~\mathrm{K}}
\right)~\mathrm{km~s}^{-1}. 
\label{zetaRT}
\end{eqnarray}
This empirical formula was derived based on the statistical distribution of 
the upper limit of $\zeta_\mathrm{RT}$, in which $V\sin I_s=0$ km s$^{-1}$
was assumed in fitting the spectral lines for a large number of stars in 
the controlled sample (the SPOCS catalog).  
Taking the ``lower" boundary of the upper limit of $\zeta_\mathrm{RT}$
as a function of $T_\mathrm{eff}$, \citet{2005ApJS..159..141V} derived
Equation (\ref{zetaRT}) \citep[see Figure 3 in][]{2005ApJS..159..141V}. 
In the subsequent analysis we assumed that the uncertainty in
$\zeta_\mathrm{RT}$ is $\pm15\%$ for cool stars ($T_\mathrm{\rm  eff}\leq6100$~K) 
based on the observed dispersion of the upper limit of $\zeta_{\rm RT}$ around
Equation (\ref{zetaRT}). 
But for hot stars ($>$6100~K) for which the SPOCS catalog has a relatively small
number of stars, we conservatively adopted $\pm25\%$ for the systematic
uncertainty in $\zeta_{\rm RT}$.

\begin{table*}[t]
\caption{
Spectroscopic Parameters. Starred systems are multiple transiting systems.
We show $\sin I_s\equiv V\sin I_s/V_\mathrm{eq}$ in the rightmost column
based on the values of $V_\mathrm{eq}$ and $V\sin I_s$. 
The listed errors in $T_\mathrm{eff}$
represent the internal statistical error and do not include the systematic
error (see Section \ref{sec:atmospheric}).
}\label{table2}
\begin{tabular}{lcccccccc}
\hline
System &$T_\mathrm{eff}$ (K) & $\log g$ & [Fe/H] 
& $M_s$ ($M_\odot$) & $R_s$ ($R_\odot$) 
& $V\sin I_s$  (km s$^{-1}$) & $V_\mathrm{eq}$ (km s$^{-1}$) &$\sin I_s$ \\\hline\hline
KOI-180 & $5592\pm 40$& $4.389 \pm 0.100$& $0.12\pm0.05$& $0.992_{-0.022}^{+0.027}$& $1.029_{-0.088}^{+0.077}$& $3.15\pm0.81$& $3.30_{-0.31}^{+0.45}$& $ 0.941_{-0.255}^{+0.276}$\\
KOI-285 & $5962\pm 27$& $3.997 \pm 0.060$& $0.16\pm0.03$& $1.324_{-0.051}^{+0.047}$& $1.914_{-0.145}^{+0.134}$& $4.21\pm0.77$& $5.74_{-0.59}^{+1.28}$& $ 0.708_{-0.162}^{+0.180}$\\
KOI-304$^\star$ & $5777\pm 42$& $4.399 \pm 0.095$& $-0.14\pm0.05$& $0.962_{-0.022}^{+0.026}$& $1.009_{-0.086}^{+0.068}$& $1.62\pm1.28$& $3.22_{-0.53}^{+0.84}$& $ 0.484_{-0.383}^{+0.425}$\\
KOI-323 & $5418\pm 30$& $4.558 \pm 0.075$& $0.01\pm0.04$& $0.927_{-0.026}^{+0.021}$& $0.841_{-0.030}^{+0.054}$& $4.70\pm0.30$& $5.56_{-0.24}^{+0.39}$& $ 0.838_{-0.071}^{+0.072}$\\
KOI-635 & $6194\pm 52$& $4.493 \pm 0.100$& $0.25\pm0.07$& $1.260_{-0.031}^{+0.026}$& $1.173_{-0.040}^{+0.050}$& $8.82\pm0.52$& $6.39_{-0.54}^{+1.08}$& $ 1.360_{-0.194}^{+0.172}$\\
KOI-678$^\star$& $5129\pm 32$& $4.532 \pm 0.085$& $0.19\pm0.04$& $0.886_{-0.018}^{+0.023}$& $0.833_{-0.046}^{+0.031}$& $3.21\pm0.45$& $3.04_{-0.18}^{+0.16}$& $ 1.058_{-0.157}^{+0.164}$\\
KOI-718$^\star$ & $6002\pm 42$& $4.577 \pm 0.080$& $0.58\pm0.04$& $1.215_{-0.022}^{+0.028}$& $1.133_{-0.044}^{+0.049}$& $2.53\pm1.26$& $3.45_{-0.31}^{+0.80}$& $ 0.697_{-0.351}^{+0.382}$\\
KOI-720$^\star$ & $5198\pm 40$& $4.580 \pm 0.100$& $0.01\pm0.05$& $0.862_{-0.020}^{+0.023}$& $0.789_{-0.038}^{+0.040}$& $4.18\pm0.30$& $4.25_{-0.20}^{+0.25}$& $ 0.980_{-0.087}^{+0.091}$\\
KOI-988$^\star$ & $5114\pm 45$& $4.544 \pm 0.115$& $0.10\pm0.04$& $0.861_{-0.019}^{+0.022}$& $0.802_{-0.045}^{+0.032}$& $2.64\pm0.57$& $3.28_{-0.19}^{+0.17}$& $ 0.808_{-0.177}^{+0.181}$\\
KOI-1615 & $5934\pm 35$& $4.266 \pm 0.080$& $0.21\pm0.04$& $1.181_{-0.038}^{+0.055}$& $1.321_{-0.135}^{+0.161}$& $8.74\pm0.21$& $8.57_{-0.91}^{+1.24}$& $ 1.017_{-0.128}^{+0.127}$\\
KOI-1628 & $6125\pm 40$& $4.274 \pm 0.085$& $0.13\pm0.04$& $1.218_{-0.033}^{+0.050}$& $1.328_{-0.137}^{+0.168}$& $11.24\pm0.27$& $11.71_{-1.42}^{+1.90}$& $ 0.957_{-0.133}^{+0.137}$\\
KOI-1779$^\star$ & $5781\pm 47$& $4.442 \pm 0.105$& $0.33\pm0.07$& $1.124_{-0.024}^{+0.027}$& $1.058_{-0.049}^{+0.133}$& $7.41\pm0.24$& $7.48_{-0.38}^{+1.03}$& $ 0.979_{-0.112}^{+0.073}$\\
KOI-1781$^\star$ & $4864\pm 55$& $4.478 \pm 0.145$& $0.19\pm0.06$& $0.815_{-0.018}^{+0.020}$& $0.760_{-0.034}^{+0.028}$& $3.64\pm0.22$& $3.67_{-0.17}^{+0.15}$& $ 0.994_{-0.072}^{+0.077}$\\
KOI-1797 & $4934\pm 42$& $4.430 \pm 0.115$& $0.16\pm0.06$& $0.824_{-0.015}^{+0.017}$& $0.781_{-0.028}^{+0.025}$& $3.69\pm0.23$& $3.65_{-0.13}^{+0.14}$& $ 1.011_{-0.072}^{+0.075}$\\
KOI-1835$^\star$ & $5046\pm 70$& $4.313 \pm 0.190$& $0.16\pm0.07$& $0.862_{-0.026}^{+0.264}$& $0.832_{-0.041}^{+1.216}$& $4.66\pm0.20$& $4.36_{-0.21}^{+6.39}$& $ 1.012_{-0.580}^{+0.126}$\\
KOI-1839 & $5465\pm 40$& $4.485 \pm 0.100$& $0.05\pm0.06$& $0.938_{-0.022}^{+0.028}$& $0.908_{-0.062}^{+0.051}$& $7.41\pm0.15$& $7.35_{-0.51}^{+0.46}$& $ 1.010_{-0.064}^{+0.077}$\\
KOI-1890 & $6107\pm 40$& $3.971 \pm 0.085$& $0.22\pm0.05$& $1.477_{-0.084}^{+0.094}$& $2.078_{-0.247}^{+0.285}$& $7.44\pm0.49$& $16.38_{-2.00}^{+2.61}$& $ 0.452_{-0.066}^{+0.073}$\\
KOI-1916$^\star$ & $5945\pm 25$& $4.308 \pm 0.060$& $0.31\pm0.04$& $1.193_{-0.025}^{+0.034}$& $1.265_{-0.094}^{+0.109}$& $6.38\pm0.39$& $6.20_{-0.52}^{+0.86}$& $ 1.017_{-0.128}^{+0.125}$\\
KOI-2001 & $5144\pm 30$& $4.484 \pm 0.085$& $0.01\pm0.04$& $0.839_{-0.014}^{+0.016}$& $0.804_{-0.029}^{+0.021}$& $2.44\pm0.66$& $2.48_{-0.10}^{+0.13}$& $ 0.980_{-0.268}^{+0.272}$\\
KOI-2002 & $5963\pm 50$& $4.071 \pm 0.110$& $0.15\pm0.05$& $1.342_{-0.080}^{+0.083}$& $1.776_{-0.263}^{+0.273}$& $5.67\pm0.42$& $8.39_{-1.30}^{+1.64}$& $ 0.671_{-0.116}^{+0.140}$\\
KOI-2026 & $5919\pm 47$& $4.178 \pm 0.100$& $0.01\pm0.05$& $1.106_{-0.041}^{+0.060}$& $1.419_{-0.175}^{+0.200}$& $4.76\pm0.50$& $7.15_{-0.98}^{+1.32}$& $ 0.659_{-0.116}^{+0.136}$\\
KOI-2035 & $5484\pm 25$& $4.544 \pm 0.060$& $0.13\pm0.04$& $0.984_{-0.023}^{+0.021}$& $0.885_{-0.024}^{+0.047}$& $6.35\pm0.21$& $6.30_{-0.22}^{+0.39}$& $ 1.002_{-0.061}^{+0.055}$\\
KOI-2087 & $5955\pm 25$& $4.364 \pm 0.060$& $0.02\pm0.03$& $1.084_{-0.020}^{+0.024}$& $1.129_{-0.075}^{+0.087}$& $4.46\pm0.73$& $4.15_{-0.47}^{+0.83}$& $ 1.048_{-0.221}^{+0.250}$\\
KOI-2261$^\star$ & $5154\pm 32$& $4.515 \pm 0.080$& $0.12\pm0.05$& $0.873_{-0.016}^{+0.021}$& $0.830_{-0.042}^{+0.027}$& $2.81\pm0.55$& $3.69_{-0.19}^{+0.17}$& $ 0.764_{-0.153}^{+0.156}$\\
KOI-2636 & $5876\pm 35$& $4.337 \pm 0.085$& $0.16\pm0.04$& $1.118_{-0.027}^{+0.037}$& $1.181_{-0.114}^{+0.145}$& $2.40\pm1.26$& $3.67_{-0.47}^{+0.84}$& $ 0.631_{-0.336}^{+0.372}$\\
\hline
\end{tabular}
\end{table*}

\subsubsection{Correction for the Impact of Differential Rotation}

The Sun's rotation period varies with surface latitude; the rotation
rate at the Sun's equator is faster than that of the polar region by
about 20\%. It is natural to assume that differential rotation is a
feature of all our program stars, and therefore that differential
rotation needs to be taken into account in our analysis.

As pointed out by \citet{2012ApJ...756...66H}, there are two main
issues that arise because of differential rotation.  The first issue
is that we do not know the latitude of the spots that are producing
the detectable photometric variations.  Starspots are probably not
randomly distributed; they are likely to be concentrated around
particular latitudes. On the Sun, the ``active latitudes'' gradually
vary from about $\pm$40$^\circ$ down to the equator, over the
11-year solar cycle. Therefore, we need to take account the systematic
errors due to the imperfect knowledge of the spots' locations. The
second issue is the distortion in the spectral line shape caused by
differential rotation. The absorption lines of a Sun-like star are
narrower than would be expected for a star with no differential
rotation, because differential rotation reduces the weight of the
extremes in rotation velocity. Therefore, an analysis of spectral
lines that neglects differential rotation will give a value of $V\sin
I_s$ that is systematically smaller than the true equatorial projected
rotation velocity.

We corrected for the first of these two issues using the procedure
described by \citet{2012ApJ...756...66H}.  Employing the empirical
relation given by \citet{2007AN....328.1030C} for the magnitude of
differential rotation, we express the rotation rate $\Omega$ as a
function of the latitude $l$ on the stellar surface:
\begin{eqnarray}
\Omega(l)=\Omega_\mathrm{eq}(1-\alpha\sin^2 l),
\label{eq:diffrot}
\end{eqnarray}
where $\Omega_\mathrm{eq}$ is the angular rotation velocity at the equator, and
\begin{eqnarray}
\alpha \Omega_\mathrm{eq}=0.053~\left( \frac{T_\mathrm{eff}}{5130~{\rm~K}}\right)^{8.6}~\mathrm{rad~day}^{-1}.
\label{eq:diffrot}
\end{eqnarray}
Assuming that the observed rotation rates are due to spots located at
the stellar latitude $l=20^\circ\pm20^\circ$ (as is the case for the
Sun), we re-estimated the equatorial rotation velocity ($V_\mathrm{eq}$)
for each of the targets as
\begin{eqnarray}
V_\mathrm{eq} = \frac{2\pi R_s}{P_s}\frac{1}{1-\alpha\sin^2 20^\circ},
\label{eq:veq}
\end{eqnarray}
and added in quadrature the following lower and
upper systematic errors in $V_\mathrm{eq}$:
\begin{eqnarray}
(\Delta V_\mathrm{eq})_\mathrm{low,sys}&=&V_\mathrm{eq}\left(\frac{1}{1-\alpha\sin^2 20^\circ}-1\right),\\
(\Delta V_\mathrm{eq})_\mathrm{upp,sys}&=&V_\mathrm{eq}
\left(\frac{1}{1-\alpha\sin^2 40^\circ}-\frac{1}{1-\alpha\sin^2 20^\circ}\right).
\label{eq:hyahha}
\end{eqnarray}
Table \ref{table2} gives the resulting estimates of the equatorial
rotation velocities.
For reference, the assumed magnitude of differential
rotation was on average $\alpha\simeq0.23$ for the targets listed in
Table \ref{table2}, which is nearly the same as that of the Sun.

Regarding the second issue, the bias in the $V\sin I_s$ measurement,
we performed a correction using the following procedure. First, we
computed $\sin I_s$ for each target based on the preliminary
measurements of $V\sin I_s$ and $V_\mathrm{eq}$ (before any correction
to $V\sin I_s$ for differential rotation). A simulated line profile
was then generated, using the model of Equation~(\ref{model}).  In
this case $M(\la)$ corresponds to the macroturbulence-plus-rotation
kernel in the presence of differential rotation using $\sin I_s$,
$V_\mathrm{eq}$, and $\alpha$ as input parameters. We adopted
plausible values for the other spectroscopic parameters (i.e., the
intrinsic Gaussian and Lorentzian dispersions, macroturbulence,
limb-darkening, and IP) in making the mock profile. This mock line was
then fitted assuming zero differential rotation, with $V\sin I_s$ as
the only free parameter. 
After computing the ratio $f$ of the resultant best-fitting $V\sin I_s$
to the product of the input $V_\mathrm{eq}$ and $\sin I_s$, we divided the originally
measured $V\sin I_s$ by the ratio $f$ to obtain the final $V\sin I_s$
corrected for the impact of differential rotation. 
We note that $f\approx1-\alpha/2$ was in general obtained, indicating that the measured 
$V\sin I_s$ is always underestimated when a rigid rotation is assumed 
in fitting the spectrum \citep[also see Figure 11 in][]{2012ApJ...756...66H}. 
The resultant $V\sin I_s$ after the correction of differential
rotation for each system is also summarized in Table \ref{table2}.

Some of our program stars were also studied by \citet{2013MNRAS.tmp.2426W},
giving us the opportunity to check on the agreement.
For the stars KOI-180, 323, and 988, respectively, 
\citet{2013MNRAS.tmp.2426W} found $V\sin I_s=2.7\pm0.5$ km s$^{-1}$, $3.3\pm0.5$ km
s$^{-1}$, and $2.7\pm0.5$ km s$^{-1}$. Comparing these with the values in
Table \ref{table2}, KOI-180 and 988 show a good agreement between two measurements, 
but KOI-323 shows a $\sim 3\sigma$ level disagreement.
Furthermore, for KOI-261, \citet{2013MNRAS.tmp.2426W} found $V\sin
I_s=2.3\pm0.5$~km~s$^{-1}$, which is in agreement with the 1$\sigma$
upper limit of 2.57~km~s$^{-1}$ determined by
\citet{2012ApJ...756...66H} using the same technique as applied here.

\section{Discussion \label{s:discussion}}\label{s:discussion}
\subsection{Evidence of Spin-orbit Misalignment \label{s:evidence}}\label{s:evidence}
\begin{figure}[t]
\begin{center}
\includegraphics[width=9cm,clip]{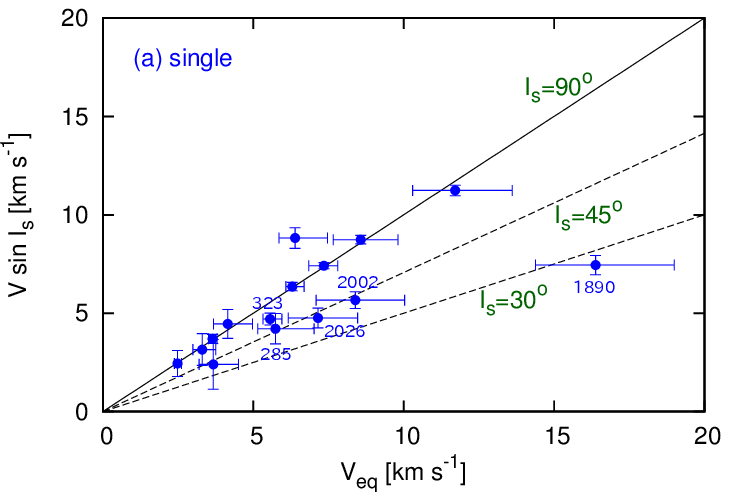} 
\includegraphics[width=9cm,clip]{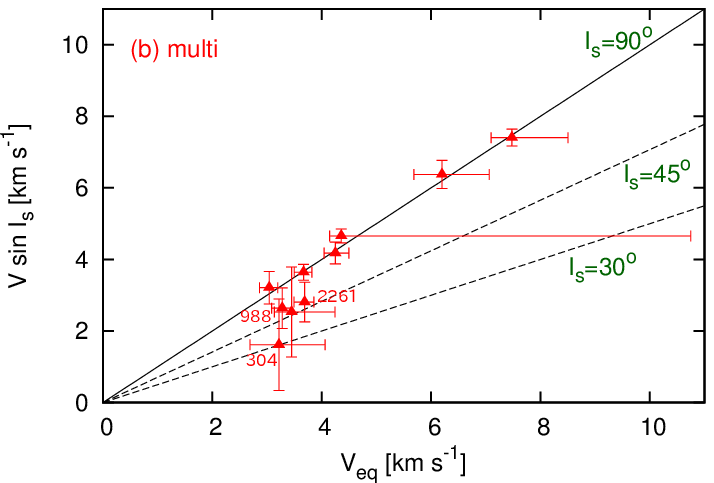} 
\caption{Projected rotational velocity ($V\sin I_s$) as a function of the stellar rotation
velocity at the equator ($V_\mathrm{eq}$), for (a) single and (b) multiple KOI systems. 
We plot here the newly observed 25 KOI systems. 
The solid lines indicates $I_s=90^\circ$ while the dashed lines 
represent different degrees of misalignment ($I_s=30^\circ$ and $I_s=45^\circ$). 
In the lower panel, the data point with the very large upper uncertainty in
$V_\mathrm{eq}$ is KOI-1835, which has a poorly determined
surface gravity (and thus a poorly determined stellar radius). 
Note that the panels (a) and (b) show different ranges of $V_\mathrm{eq}$.
}\label{fig:v_vsini}
\end{center}
\end{figure}
Figure \ref{fig:v_vsini} plots $V\sin I_s$ against $V_\mathrm{eq}$, after
making the corrections for differential rotation. 
Single transiting systems are shown in panel (a),
and systems with multiple transiting candidates are shown in panel (b).
The black solid line represents $I_s=90^\circ$.
Systems falling on this line would have 
the stellar spin oriented perpendicular to
the line-of-sight, and therefore likely aligned with the planetary orbital axes
(although an unlikely possibility is that they are misaligned with the line of
nodes coincidentally along the line of sight).
The dashed lines show different degrees of misalignment
($I_s=45^\circ$ and $I_s=30^\circ$). 

Most of the data points in Figure \ref{fig:v_vsini} do indeed fall
near the $I_s=90^\circ$ line, indicating a tendency toward spin-orbit
alignment. Four of the systems---KOI-323, 1890, 2002, and 2026---show
evidence for significant spin-orbit misalignments with more than
$2\sigma$ confidence. All four of these systems are single-transiting
candidates.  Some of the multiple-transiting candidates also show
evidence for misalignment but only at the 1$\sigma$ level; these are
KOI-304, 988, and 2261.

\begin{figure}[t]
\begin{center}
\includegraphics[width=8.8cm]{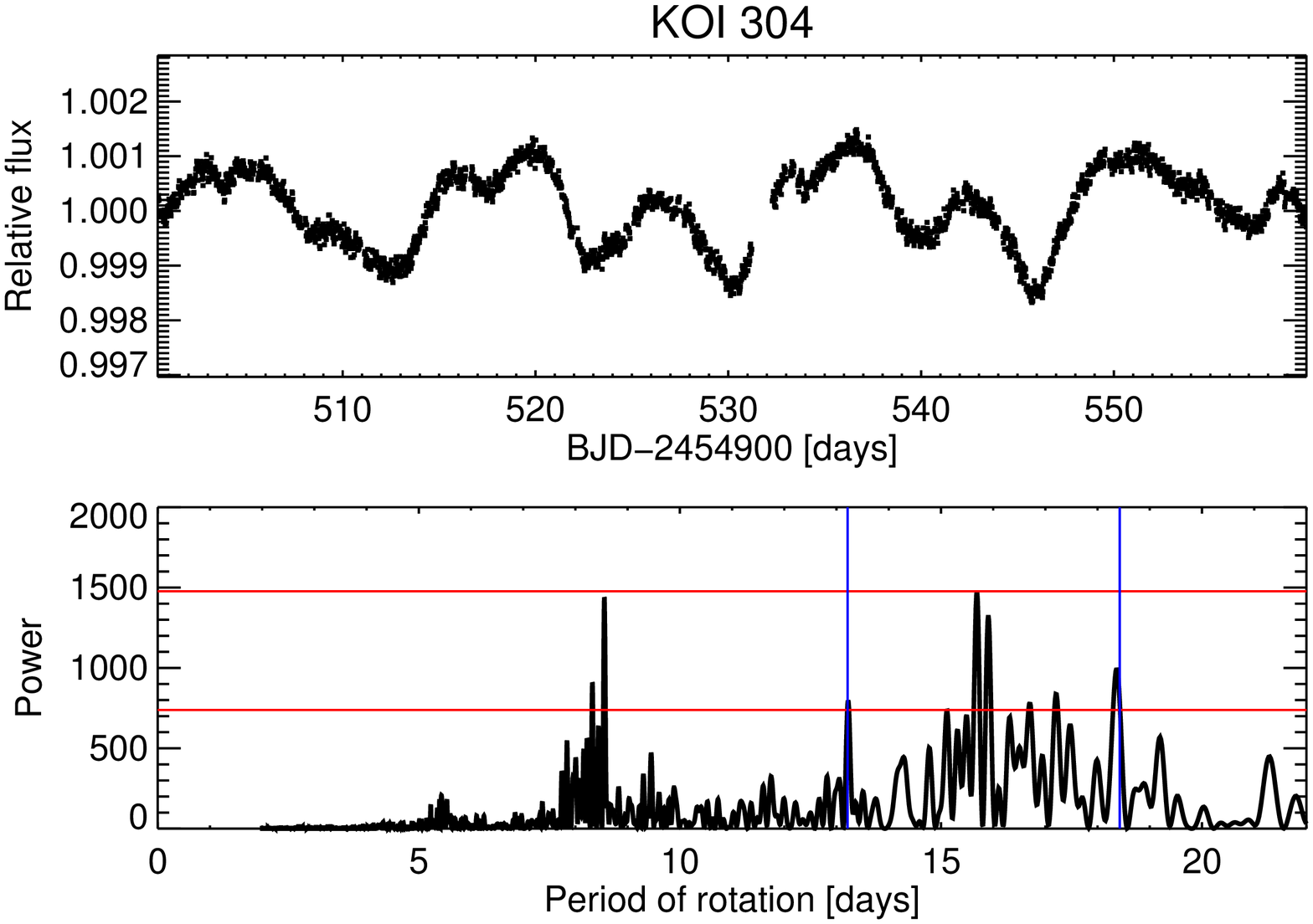}
\includegraphics[width=8.8cm]{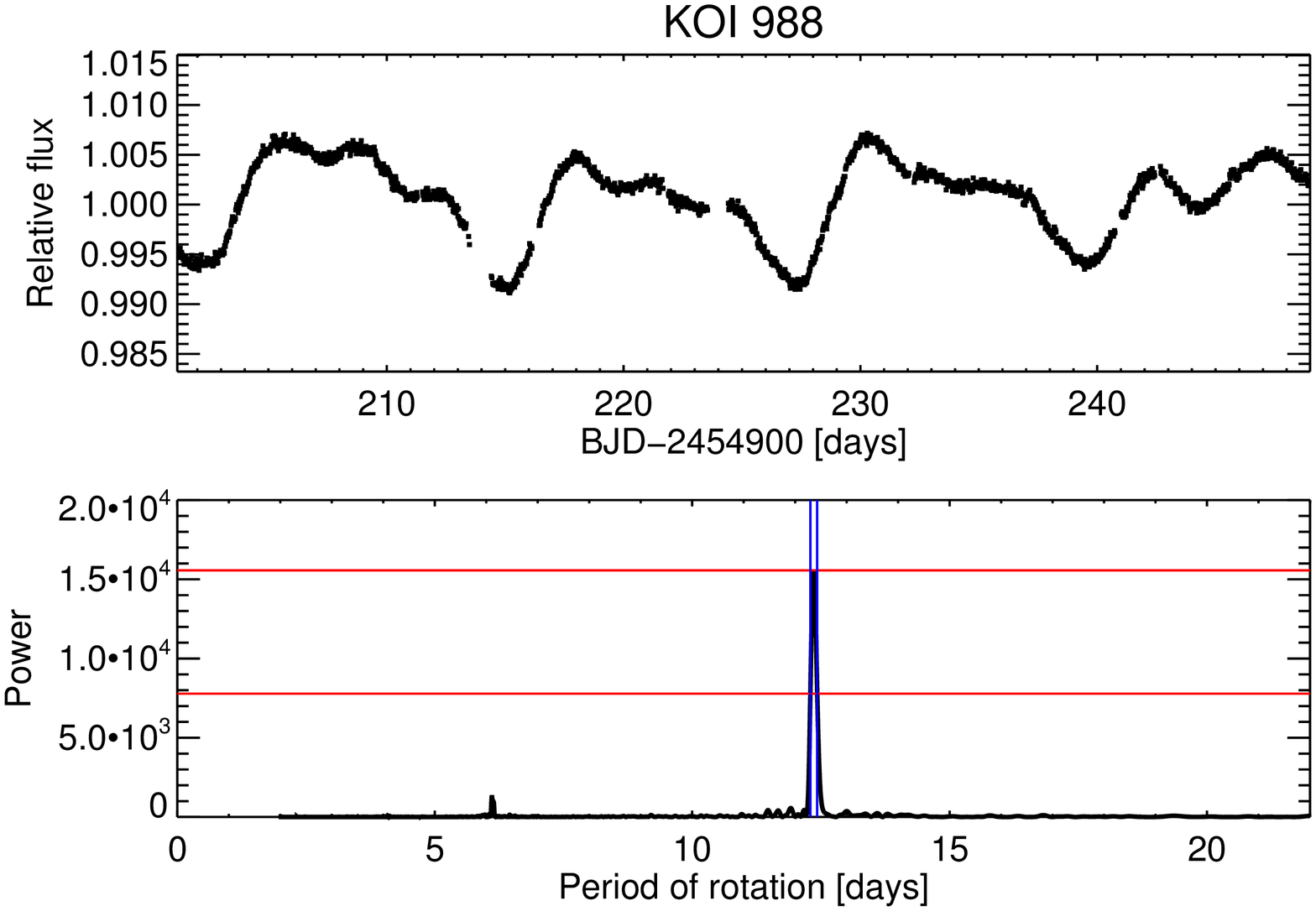}
\includegraphics[width=8.8cm]{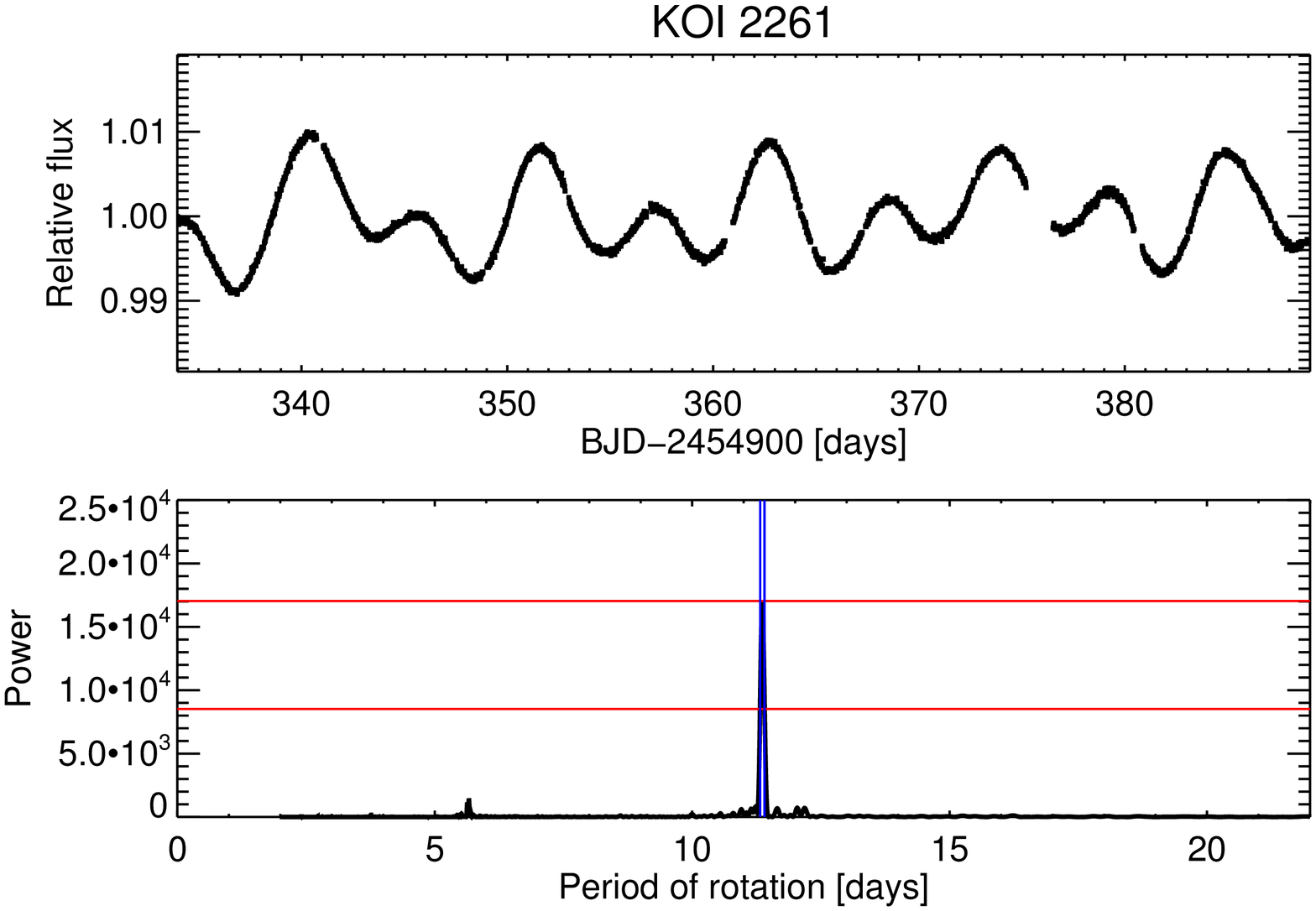}
\caption{Light curves and Lomb-Scargle periodograms for multiple systems showing
a possible spin-orbit misalignment (KOI-304, 988, 2261). 
In the periodograms, the intervals surrounded by the two blue lines correspond to
the rotation periods and their uncertainties (\S \ref{s:periods_estimate}). 
}\label{fig:periodogram}
\end{center}
\end{figure}
As the multiple-transiting systems are of special importance, it is
worth focusing on those possible misalignments and check if the
results for the rotation period, stellar radius, and $V\sin I_s$ are
robust. A spurious finding of misalignment can result from an underestimate
of either $V\sin I_s$ or $P$, or an overestimate of $R_s$.

\begin{figure}[t]
\begin{center}
\includegraphics[width=8.8cm]{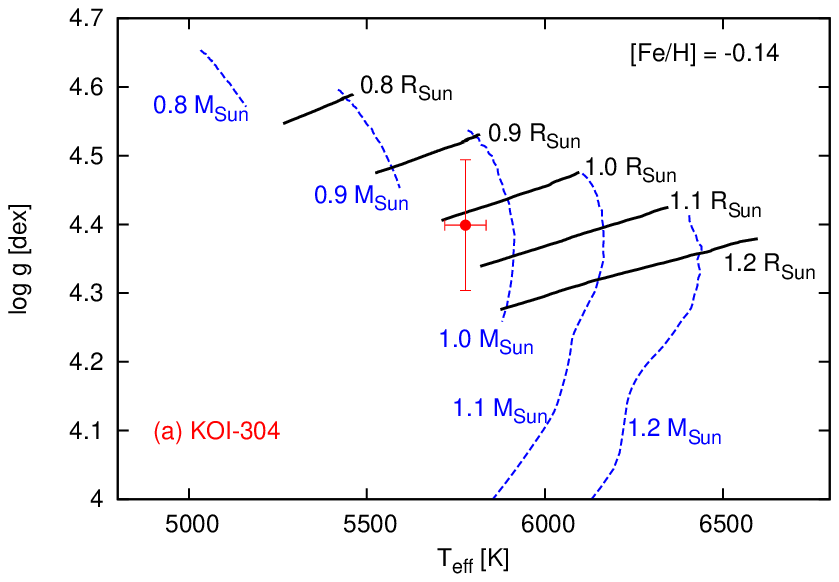}
\includegraphics[width=8.8cm]{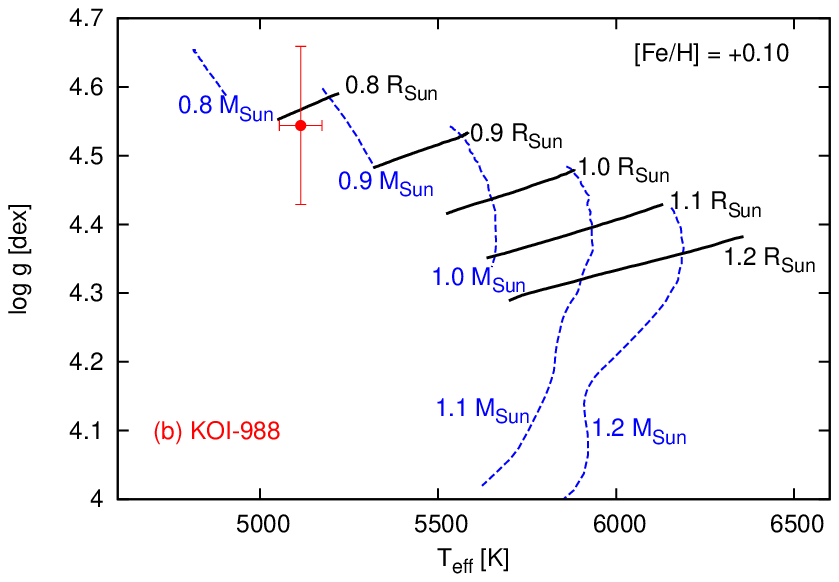}
\includegraphics[width=8.8cm]{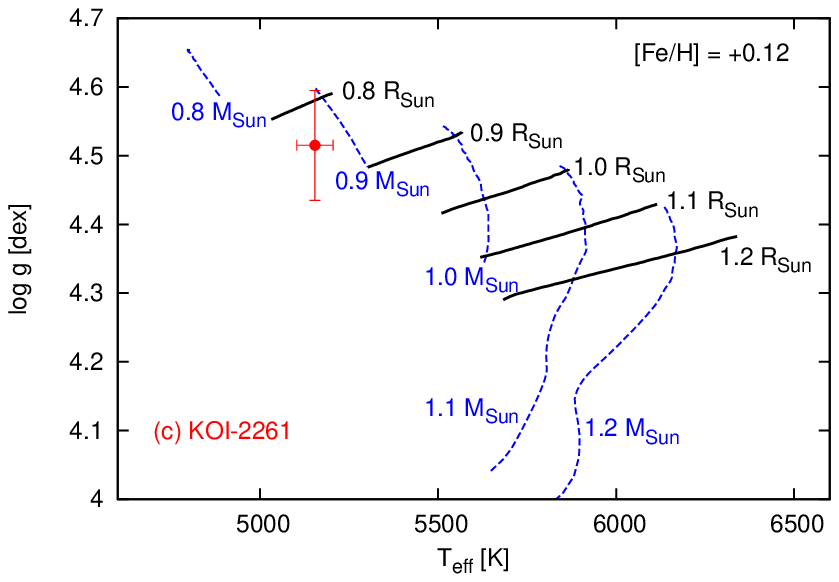}
\caption{Placements of measured $T_\mathrm{eff}$ and $\log g$ in the Y$^2$ isochrone
for (a) KOI-304, (b) KOI-988, and (c) KOI-2261 (red crosses with error bars). 
The blue dashed lines indicate the evolutional tracks and
the black solid lines are the ``iso-radius", based on the Y$^2$ isochrone. 
Note that the errorbars in $T_\mathrm{eff}$ are enlarged so that they contain
the systematic error of $40$ K.
}\label{fig:isochrone}
\end{center}
\end{figure}
First, we check on the rotation periods. The relevant light curves and
periodograms are shown in Figure \ref{fig:periodogram}.  Each light
curve shown in Figure \ref{fig:periodogram} shows an evident pattern of quasi-periodic flux
variation, and the periodograms for KOI-988 and KOI-2261 exhibit a
clear and unambiguous peak that surpasses a power of $10^4$.  For the
case of KOI-304, on the other hand, there are multiple,
relatively weak peaks of comparable power.  These multiple peaks could
be ascribed to differential rotation or rapid starspot evolution.
Nevertheless, visual inspection of the light curves does not reveal
any problem with the quoted rotation periods of $15.8\pm 2.6$~days for KOI-304. 

Next, we check on the determination of the stellar radius.  We have
already shown in Section \ref{sec:atmospheric} that the photospheric
parameters ($T_\mathrm{eff}$ and $\log g$) are in reasonably good
agreement with the KIC values. Here we focus on the estimate of
stellar mass and radius, based on the Y$^2$ isochrones.  Figure
\ref{fig:isochrone} shows the placement of the measured values of
$T_\mathrm{eff}$ and $\log g$ (red crosses) on the theoretical
isochrones (blue dashed lines) and the loci of equal stellar radius
(black solid lines) of the Y$^2$ theoretical evolutionary models for
main-sequence stars.  The measured values of $T_\mathrm{eff}$ and
$\log g$ for KOI-304, 988, and 2261 conform with the models. 

\begin{figure}[t]
\begin{center}
\includegraphics[width=8.8cm]{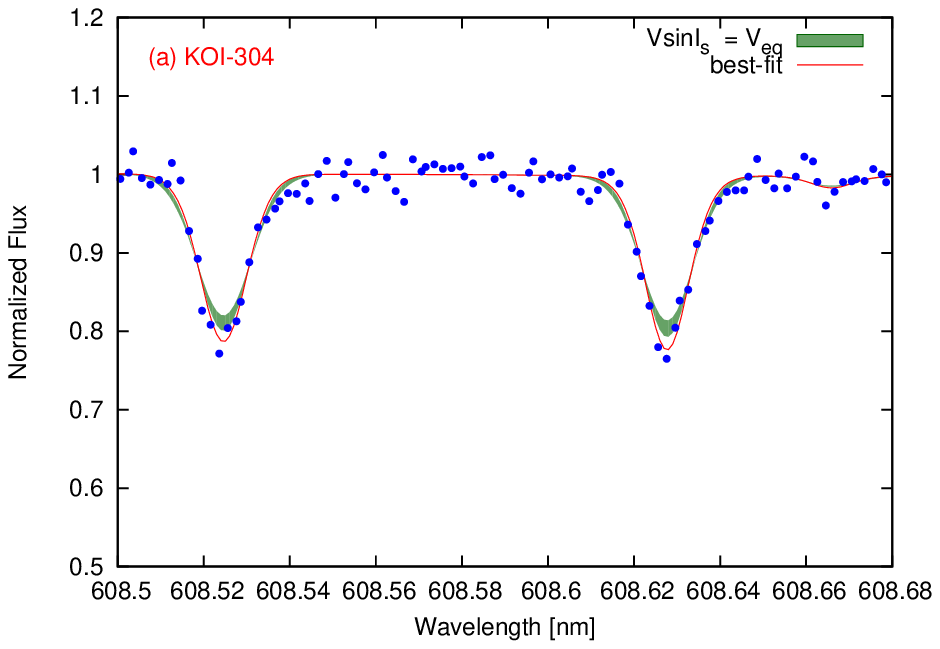}
\includegraphics[width=8.8cm]{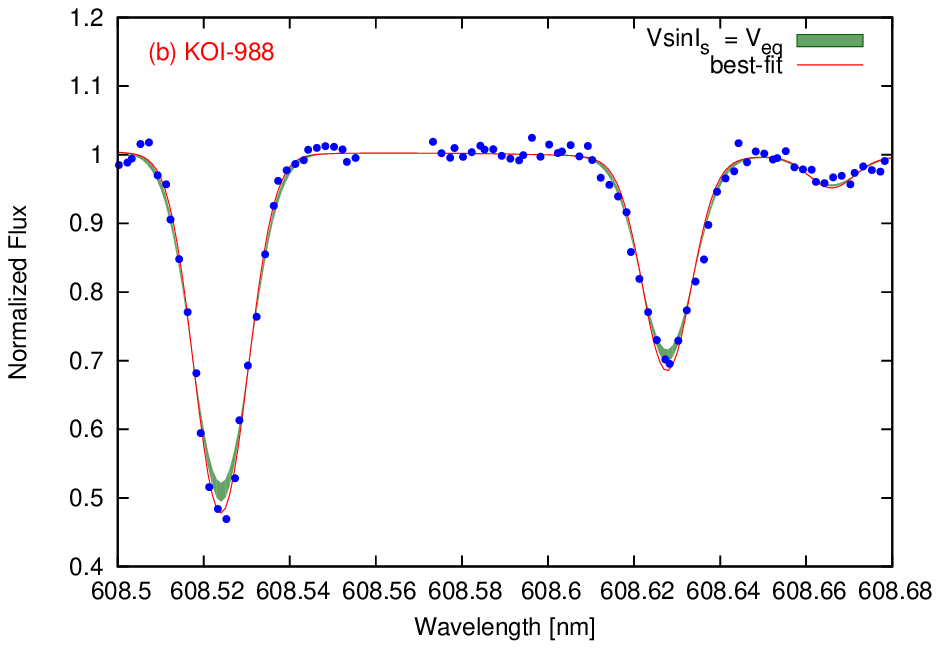}
\includegraphics[width=8.8cm]{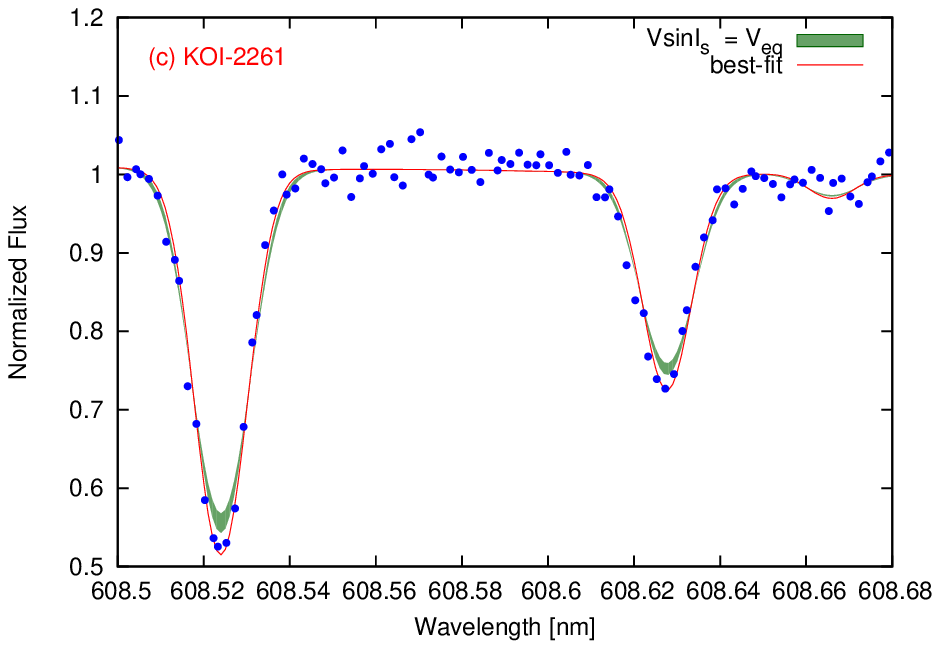}
\caption{Part of the spectrum used for fitting $V\sin I_s$ for (a) KOI-304, (b) KOI-988, and (c) KOI-2261. 
In each panel, the blue dots show the observed spectrum and the red solid line indicates the best-fitting model
computed by Equation (\ref{model}). 
The green area indicates the range of model spectra satisfying $\sin I_s=1$ (i.e., spin-orbit alignment)
with a macroturbulent velocity $\zeta_\mathrm{RT}$ differing by $\pm$15\%
from the assumed value.
}\label{fig:spec}
\end{center}
\end{figure}
Finally, we check on the measurements of $V\sin I_s$ based on the
observed line broadening in the Subaru spectra.
Figure \ref{fig:spec} shows part of the observed spectrum (blue dots) along with 
the best-fitting model spectrum (red line) for each of (a) KOI-304, (b) KOI-988, and (c) KOI-2261.
For reference, the green area shows the spectral lines that would be
expected for $\sin I_s=1$ (i.e., spin-orbit alignment). The breadth of
the green area arises from variation of the
macroturbulent velocity $\zeta_\mathrm{RT}$ by $\pm$15\%
from the value computed by Equation (\ref{zetaRT}).
A misaligned system will show narrower lines than the green region.
This figure illustrates the main difficulty of this probe of
spin-orbit alignment: one must isolate the very small differences in
line broadening due to rotation as opposed to macroturbulence and
instrumental broadening. 
For all the three systems shown here, $V\sin I_s$ cannot be much larger
than the values listed in Table \ref{table2} (see in particular the
bottom of each absorption line), unless the assumed macroturbulent
velocity is in error by more than 15\%.
It is important to remember that Figure \ref{fig:spec} shows only a
part of the observed spectrum. The true statistical significance of
the results is higher than it might seem visually because $V\sin I_s$
was determined from data over a wider range of wavelengths.

In summary, the detailed visual inspection of the multi-transiting
systems with possible misalignments 
did not raise any specific concerns for all of KOI-304, 988, and 2261, 
which remain viable candidates for multi-planet systems with
misaligned stars. Each individual detection is statistically marginal,
with less than 2$\sigma$ confidence, but if the uncertainties have
been accurately determined, then together it is likely that at least
one system is misaligned.  To quantify this statement we can compute
the probability that all three multiple systems (KOI-304, 988, 2261)
are well-aligned, defining this for convenience to mean $I_s \geq
75^\circ$ ($\sin I_s \geq 0.9659$). Assuming that both $V_\mathrm{eq}$
and $V\sin I_s$ have uncertainties drawn from independent Gaussian
distributions, with dispersions set equal to our quoted uncertainties,
we calculate the probability for each system to have $0.9659\leq \sin I_s$. 
If the lower and upper observation errors are different, we
adopt a two-sided Gaussian with different upper and lower
dispersions. We then compute the products of the resulting
probabilities to find the net probability $p_\mathrm{all~aligned}$
that all three systems are aligned.  We find
$p_\mathrm{all~aligned}=0.0025$,
implying that at least one system
among the three KOI's is very likely to have spin-orbit misalignment.
This result cannot be definitive, though, given the possibility of
systematic effects, or uncertainties that are correlated between
different systems due to shared assumptions and techniques. 
Specifically, all the three systems fall on the regime where
the measurement of $V\sin I_s$ tends to suffer from systematic effects
($V\sin I_s\lesssim 3$ km s$^{-1}$).
It is better to regard KOI-304, 988, and 2261 as candidate misalignments
that are good targets for additional follow-up observations.


\subsection{Distribution of Stellar Inclinations}

In the previous subsection, we have seen that some of the systems
(both single and multiple) may have spin-orbit misalignments.  A
natural question is ``what is the fraction of misaligned systems?''
Although the number of our samples is still small, a histogram of the
observed $I_s$ may be helpful to gain an insight into the underlying
true distribution of the spin-orbit angle, just as the histogram of
sky-plane angles was useful in the case of RM measurements
\citep[e.g.,][]{2010MNRAS.402L...1P}.  One issue concerning the
conversion from the observed $V_\mathrm{eq}$ and $V\sin I_s$ to the
distribution of $I_s$ is that $\sin I_s\equiv V\sin I_s/V_\mathrm{eq}$
could extend beyond unity due to measurement uncertainties.
Theoretical distributions of $\sin I_s$ always satisfy $0\leq\sin
I_s\leq 1$, which inhibits a direct comparison between the theoretical
and observed distributions.  Here, we present a Bayesian method that
avoids this problem by placing a prior on $I_s$.

Based on Bayes' theorem, the posterior probability distribution 
of $V_\mathrm{eq}$ and $I_s$ is 
\begin{eqnarray}
P\left(V_\mathrm{eq}, I_s  | \mathrm{D}\right)
\propto P(\mathrm{D} | V_\mathrm{eq}, I_s)
\cdot p_\mathrm{prior}(V_\mathrm{eq})\cdot p_\mathrm{prior}(I_s),
\end{eqnarray}
where ``D" represents the observed data
for $V_\mathrm{eq}$ and $V\sin I_s$ for each of the observed systems. 
We again assume that observational data for $V_\mathrm{eq}$ and $V\sin I_s$ follow
the Gaussian distributions with their centers being $V_\mathrm{eq}=p^{(i)}$ and $V\sin I_s=q^{(i)}$,
and dispersions being $\sigma_p^{(i)}$ and $\sigma_q^{(i)}$, where $i$ is the label of the system. 
In this case, the conditional probability $P(\mathrm{D} | V_\mathrm{eq}, I_s)$
is expressed as
\begin{eqnarray}
P(\mathrm{D} | V_\mathrm{eq}, I_s)
\propto 
\frac{1}{\sigma_p^{(i)}}\exp\left\{-\frac{(p^{(i)}-V_\mathrm{eq})^2}{2\sigma_p^{(i)2}}\right\}\nonumber\\
\frac{1}{\sigma_q^{(i)}}\exp\left\{-\frac{(q^{(i)}-V_\mathrm{eq}\sin I_s)^2}{2\sigma_q^{(i)2}}\right\}. 
\end{eqnarray}
Assuming a uniform distribution for $p_\mathrm{prior}(V_\mathrm{eq})$ ($0\leq V_\mathrm{eq}$), 
we marginalize $V_\mathrm{eq}$, so that we obtain the posterior distribution for $I_s$:
\begin{eqnarray}
P_i(I_s  | \mathrm{D})
\propto \int_0^\infty \frac{1}{\sigma_p^{(i)}}\exp\left\{-\frac{(p^{(i)}-V_\mathrm{eq})^2}{2\sigma_p^{(i)2}}\right\}\nonumber\\
\frac{1}{\sigma_q^{(i)}}\exp\left\{-\frac{(q^{(i)}-V_\mathrm{eq}\sin I_s)^2}{2\sigma_q^{(i)2}}\right\}dV_\mathrm{eq}
\cdot p_\mathrm{prior}(I_s).
\label{eq:posterior}
\end{eqnarray}
In case that the observed result for $V_\mathrm{eq}$ or $V\sin I_s$ has different upper and lower errors, 
we adopt two-sided Gaussian functions as in \S \ref{s:evidence}. 
When a prior defined in $0\leq I_s\leq \pi/2$ (i.e., $0\leq\sin I_s\leq1$) is applied,
the posterior $P_i(I_s  | \mathrm{D})$ also could have a non-zero value
in $0\leq I_s\leq \pi/2$. 

\begin{figure}[t]
\begin{center}
\includegraphics[width=9cm,clip]{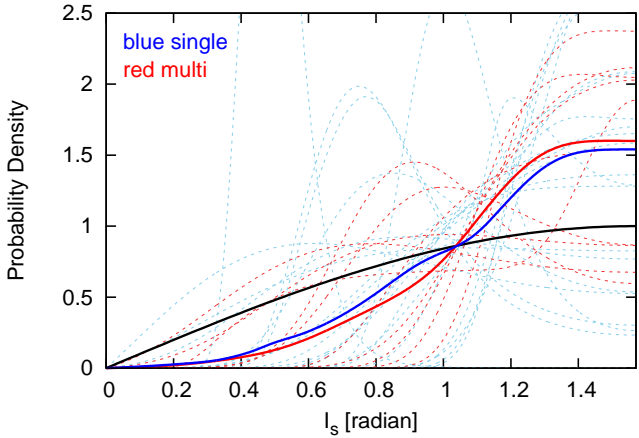} 
\caption{
Posterior distributions of $I_s$ computed by Equation (\ref{eq:posterior}).
The dashed lines correspond to the result for each system in the sample
(light-blue for single and light-red for multiple systems). 
The solid lines are averaged shapes of the posteriors
for each category. The black solid curve
represents the isotropic distribution of $I_s$ for reference. 
}\label{fig:posterior}
\end{center}
\end{figure}
For each of the observed KOI systems, we compute the posterior 
distribution by Equation (\ref{eq:posterior}). 
We here assume the isotropic distribution for the prior on $I_s$
(i.e., $p_\mathrm{prior}(I_s)=\sin I_s$). This is physically unlikely
considering the fact that many systems show a good spin-orbit alignment
from measurements of the RM effect. 
However, the prior distribution is not so important since we do not attempt to quantitatively
compare any distributions here (see the next subsection for a quantitative 
comparison). Instead, in order to visualize the distribution of $I_s$, we take the average
of the posterior distributions by stacking $P_i(I_s  | \mathrm{D})$ for
observed systems. In Figure \ref{fig:posterior}, we plot the averaged posterior
distribution for either of single (blue) and multiple (red) KOI systems
by the solid lines. 
These plots correspond to a sort of histogram of $I_s$ considering that
the peak of the posterior for each system likely represents the 
most plausible value of $I_s$, and all the systems have an equal weight. 
The two distributions (single and multiple) are similar, but 
single systems show a slightly wider distribution than that of
multiple systems with small bumps at $I_s\lesssim1.0$ radian ($\simeq 57^\circ$). 
For reference, we show by the black solid line the isotropic distribution of $I_s$. 
Note that in this analysis (and other statistical analyses below), we added
the seven KOI systems reported in our previous campaign 
\citep[KOI-257, 261, 262, 269, 280, 367, 974,][]{2012ApJ...756...66H} to the list of 
targets subjected to the statistical analysis. 

It should be stressed that our observed systems (both single and multiple) 
have no hot Jupiters and all the planet candidates are Earth-sized or 
Neptune-sized ones. Little is known about the spin-orbit angle for these classes
of planets, and the observed distribution of the angle could
be more or less different from that for close-in giant planets. 
We also note that while hot Jupiters are in general isolated single planets \citep{2012PNAS..109.7982S}, 
many of single transiting systems in our sample may actually be multiple systems
(e.g., transits of outer planets are unobservable due to geometry). 
This possibility makes it difficult to interpret the comparison 
of $I_s$ for single and multiple systems.

\subsection{Statistical Tests\label{s:test}}\label{s:test}

Figures \ref{fig:v_vsini} and \ref{fig:posterior} suggest that the
observed distributions of $I_s$ differ from an isotropic distribution
and also differ from perfect spin-orbit alignment.  However, the
degree to which the observed distributions are different from or
similar to each other is quantitatively not clear.  Also, we are
interested in whether single-transiting and multiple-transiting
systems have the same distribution for $I_s$. We test the following
two hypotheses with the Kolmogorov-Smirnov (KS) test:
\begin{enumerate}
\item[(a)] the observed values of $I_s$ (all systems) are drawn from an isotropic distribution, 
\item[(b)] the observed distributions of $I_s$ for single and multiple systems are
the same. 
\end{enumerate}

We perform a Monte-Carlo simulation to 
implement the KS tests. We take the following steps based on the observed
values of $V_\mathrm{eq}$ and $V\sin I_s$.
\begin{enumerate}
\item First, we randomly generate $(V_\mathrm{eq}^{(i)}, V\sin I_s^{(i)})$ 
for system $i$ assuming Gaussian distributions (two-sided Gaussians if needed)
with dispersions set equal to the quoted measurement uncertainties.
\item We then compute $I_s^{(i)}\equiv\arcsin(V\sin I_s^{(i)}/V_\mathrm{eq}^{(i)})$
for each system.
Whenever $V\sin I_s^{(i)}>V_\mathrm{eq}^{(i)}$, we set $I_s=90^\circ$. 
\item Based on the set of \{$I_s^{(i)}$\} with all the systems in the sample, 
we implement the KS test and record the value of $D$ (the largest difference between the two cumulative distributions). 
\item We repeat the preceding steps (1 to 3) $10^6$ times, 
recording the values of $D$ and finding the median and standard deviation
of the collection of $D$ values, and the corresponding probability that the two distributions 
may be the same, which we denote by $p(D>D_\mathrm{obs})$. 
\end{enumerate}

We first test the hypothesis (a). The two distributions tested are the observed
distribution of \{$I_s^{(i)}$\} and the theoretical isotropic distribution. As a result
of implementing the steps 1 - 4, we obtain 
$D_\mathrm{obs}=0.344_{-0.063}^{+0.094}$,
corresponding to $p(D>D_\mathrm{obs})=0.00071_{-0.00071}^{+0.00912}$.
Therefore, the hypothesis (a) is highly unlikely, and this result should indicate
that the stellar equators in our sample are preferentially edge-on, suggesting
a tendency toward spin-orbit alignment. 

In the second test, the two tested distributions are both observed distributions
of $I_s$, one for single and the other for multiple systems. Implementing the 
KS test, we find 
$D_\mathrm{obs}=0.255_{-0.065}^{+0.091}$, which corresponds to 
$p(D>D_\mathrm{obs})=0.665_{-0.380}^{+0.265}$.
This result indicates that the two observed distributions are not significantly different, 
and might be drawn from the same distribution. 
To see this result is robust, we repeat the steps 1-4, implementing instead of the KS test 
the $K$-sample Anderson-Darling (AD) test \citep[e.g.,][]{2009ApJ...702.1199H}, which 
has more sensitivity around the tails of distributions. 
As a consequence, the $p$-value of 
$0.519_{-0.277}^{+0.123}$
is obtained, which also implies the two observed distributions are not
significantly different.  The results of these statistical tests
cannot corroborate recent findings by RM measurements,
asteroseismology, and the spot-crossing method that
multiple-transiting systems preferentially show a good spin-orbit
alignment \citep{2012Natur.487..449S, 2012ApJ...759L..36H,
  2013ApJ...771...11A, 2013ApJ...766..101C}.  However, it is premature
to conclude that our result actually contradicts the previous
findings; more systems are needed (particularly multiple-transiting
systems) for a more definitive conclusion.  The exact sample size that
will be required depends on the true distribution of the spin-orbit
angle.

\section{Summary \label{s:summary}}\label{s:summary}

In this paper, we investigated the stellar inclinations for KOI 
systems by combining the rotation periods estimated from the \textit{Kepler} 
photometry and projected rotational velocities $V\sin I_s$
determined from Subaru spectroscopy. We constrained the stellar inclination
$I_s$ for 25 KOI systems, and discussed statistical properties using
all the systems observed so far by Subaru. There are several implications 
that we list here.
\begin{enumerate}
\item Based on the KS test, the observed distribution of $I_s$ 
is significantly different from an isotropic distribution, suggesting 
that the direction of stellar spin is correlated with
the planetary orbital axis. Spin-orbit alignment has been reported for 
many transiting systems, but most of the systems with
RM measurements have hot (warm) Jupiters.
Our measurements pertain to Neptune-sized or Earth-sized planets,
which are likely to have a different history of formation
and migration than giant planets.
In particular the smaller planets are not
as likely to have strong tidal interactions with their host stars,
and therefore the orbital orientations may reflect
more primordial conditions.
\item 
A certain fraction of the systems show possible spin-orbit 
misalignments ($I_s\lesssim 75^\circ$). 
We had a closer look at the 
seemingly misaligned multiple transiting systems (KOI-304, 988, and 2261), 
and they all survived as candidates for misaligned stars.
\item The statistical tests indicate that the observed distributions of $I_s$
for single and multiple transiting systems are not significantly different. 
The sensitivity of this test is limited, however, by the small number
of multiple systems (only 11).
The averaged posterior distribution
shown in Figure \ref{fig:posterior} suggests that the single transiting
systems might have a larger fraction of spin-orbit misalignment. 
This should be confirmed or refuted by further observations of
transiting systems. 
\end{enumerate}

As \citet{2012ApJ...756...66H} noted, our present method cannot discriminate the state of 
$I_s=-90^\circ$ (retrograde orbit) from that of $I_s=+90^\circ$ (prograde orbit). 
The degeneracy between $I_s = -90^\circ$ and $I_s = +90^\circ$ would certainly make the fraction 
of misaligned systems look smaller than the real fraction (systems with $I_s \approx -90^\circ$ 
would appear to be aligned in Figure \ref{fig:v_vsini}), but it does not affect the statements 1. and 2. 
of the above summary. In addition, given the fact that the measurements of the RM effect 
so far have not revealed a strong evidence of a ``perfectly anti-aligned" system 
(i.e., $\la \approx \pm 180^\circ$), it is expected to be a rare case to find a system with $I_s \approx -90^\circ$. 
All the other retrograde cases (e.g., $-75^\circ \lesssim I_s \lesssim 0^\circ$) are actually regarded 
as "misaligned" in Figure \ref{fig:v_vsini} as in the case of prograde orbits. 
In other words, our methodology gives the lower limit on the fraction of misaligned systems.

One task left is the confirmation of the planetary nature for the KOI
planet candidates on which we focused in this paper.  While the false
positive rate for KOI multiple systems is proved to be negligible
\citep{2012ApJ...750..112L}, any contamination from
background/foreground source(s) leads to a wrong determination of the
rotation period and/or spectroscopic parameters. A deep direct imaging
search for companions around the KOI stars would be helpful both in
terms of putting a constraint on the magnitude of contamination and
identifying the possible cause of spin-orbit misalignment.

\acknowledgments 

This paper is based on data collected at Subaru Telescope, which is
operated by the National Astronomical Observatory of Japan.  
We acknowledge the support for our Subaru HDS observations by Akito
Tajitsu, a support scientist for the Subaru HDS.  T.H.\  
expresses special thanks to Masayuki Kuzuhara, Yuka Fujii, Akihiko Fukui, and Yasushi Suto 
for fruitful discussions on this subject. The data analysis was in part carried out on common 
use data analysis computer system at the Astronomy Data Center, ADC, of the
National Astronomical Observatory of Japan.  
T.H.\ and Y.H.T.\ are supported by Japan Society for Promotion of Science (JSPS) 
Fellowship for Research (PD:25-3183, DC1: 23-3491).  
J.N.W.\ and R.S.O.\ gratefully acknowledge support from the NASA Origins program
(NNX11AG85G)
and Kepler Participating Scientist program (NNX12AC76G).
N.N.\ acknowledges support by the NAOJ Fellowship, the NINS Program
for Cross-Disciplinary Study, and Grant-in-Aid for Scientific
Research (A) (No. 25247026) from the Ministry of Education, Culture, Sports, 
Science and Technology (MEXT) of Japan.
We acknowledge the very significant cultural role and reverence that the
summit of Mauna Kea has always had within the indigenous people in Hawai'i. 



\begin{thebibliography}{54}
\expandafter\ifx\csname natexlab\endcsname\relax\def\natexlab#1{#1}\fi

\bibitem[{{Akeson} {et~al.}(2013){Akeson}, {Chen}, {Ciardi}, {Crane}, {Good},
  {Harbut}, {Jackson}, {Kane}, {Laity}, {Leifer}, {Lynn}, {McElroy}, {Papin},
  {Plavchan}, {Ram{\'{\i}}rez}, {Rey}, {von Braun}, {Wittman}, {Abajian},
  {Ali}, {Beichman}, {Beekley}, {Berriman}, {Berukoff}, {Bryden}, {Chan},
  {Groom}, {Lau}, {Payne}, {Regelson}, {Saucedo}, {Schmitz}, {Stauffer},
  {Wyatt}, \& {Zhang}}]{2013PASP..125..989A}
{Akeson}, R.~L., {et~al.} 2013, \pasp, 125, 989

\bibitem[{{Albrecht} {et~al.}(2013){Albrecht}, {Winn}, {Marcy}, {Howard},
  {Isaacson}, \& {Johnson}}]{2013ApJ...771...11A}
{Albrecht}, S., {Winn}, J.~N., {Marcy}, G.~W., {Howard}, A.~W., {Isaacson}, H.,
  \& {Johnson}, J.~A. 2013, \apj, 771, 11

\bibitem[{{Albrecht} {et~al.}(2012){Albrecht}, {Winn}, {Johnson}, {Howard},
  {Marcy}, {Butler}, {Arriagada}, {Crane}, {Shectman}, {Thompson}, {Hirano},
  {Bakos}, \& {Hartman}}]{2012ApJ...757...18A}
{Albrecht}, S., {et~al.} 2012, \apj, 757, 18

\bibitem[{{Batalha} {et~al.}(2013){Batalha}, {Rowe}, {Bryson}, {Barclay},
  {Burke}, {Caldwell}, {Christiansen}, {Mullally}, {Thompson}, {Brown},
  {Dupree}, {Fabrycky}, {Ford}, {Fortney}, {Gilliland}, {Isaacson}, {Latham},
  {Marcy}, {Quinn}, {Ragozzine}, {Shporer}, {Borucki}, {Ciardi}, {Gautier},
  {Haas}, {Jenkins}, {Koch}, {Lissauer}, {Rapin}, {Basri}, {Boss}, {Buchhave},
  {Carter}, {Charbonneau}, {Christensen-Dalsgaard}, {Clarke}, {Cochran},
  {Demory}, {Desert}, {Devore}, {Doyle}, {Esquerdo}, {Everett}, {Fressin},
  {Geary}, {Girouard}, {Gould}, {Hall}, {Holman}, {Howard}, {Howell},
  {Ibrahim}, {Kinemuchi}, {Kjeldsen}, {Klaus}, {Li}, {Lucas}, {Meibom},
  {Morris}, {Pr{\v s}a}, {Quintana}, {Sanderfer}, {Sasselov}, {Seader},
  {Smith}, {Steffen}, {Still}, {Stumpe}, {Tarter}, {Tenenbaum}, {Torres},
  {Twicken}, {Uddin}, {Van Cleve}, {Walkowicz}, \&
  {Welsh}}]{2013ApJS..204...24B}
{Batalha}, N.~M., {et~al.} 2013, \apjs, 204, 24

\bibitem[{{Borucki} {et~al.}(2010){Borucki}, {Koch}, {Basri}, {Batalha},
  {Brown}, {Caldwell}, {Caldwell}, {Christensen-Dalsgaard}, {Cochran},
  {DeVore}, {Dunham}, {Dupree}, {Gautier}, {Geary}, {Gilliland}, {Gould},
  {Howell}, {Jenkins}, {Kondo}, {Latham}, {Marcy}, {Meibom}, {Kjeldsen},
  {Lissauer}, {Monet}, {Morrison}, {Sasselov}, {Tarter}, {Boss}, {Brownlee},
  {Owen}, {Buzasi}, {Charbonneau}, {Doyle}, {Fortney}, {Ford}, {Holman},
  {Seager}, {Steffen}, {Welsh}, {Rowe}, {Anderson}, {Buchhave}, {Ciardi},
  {Walkowicz}, {Sherry}, {Horch}, {Isaacson}, {Everett}, {Fischer}, {Torres},
  {Johnson}, {Endl}, {MacQueen}, {Bryson}, {Dotson}, {Haas}, {Kolodziejczak},
  {Van Cleve}, {Chandrasekaran}, {Twicken}, {Quintana}, {Clarke}, {Allen},
  {Li}, {Wu}, {Tenenbaum}, {Verner}, {Bruhweiler}, {Barnes}, \&
  {Prsa}}]{2010Sci...327..977B}
{Borucki}, W.~J., {et~al.} 2010, Science, 327, 977

\bibitem[{{Brown} {et~al.}(2011){Brown}, {Latham}, {Everett}, \&
  {Esquerdo}}]{2011AJ....142..112B}
{Brown}, T.~M., {Latham}, D.~W., {Everett}, M.~E., \& {Esquerdo}, G.~A. 2011,
  \aj, 142, 112

\bibitem[{{Bruntt} {et~al.}(2010){Bruntt}, {Bedding}, {Quirion}, {Lo Curto},
  {Carrier}, {Smalley}, {Dall}, {Arentoft}, {Bazot}, \&
  {Butler}}]{2010MNRAS.405.1907B}
{Bruntt}, H., {et~al.} 2010, \mnras, 405, 1907

\bibitem[{{Chaplin} {et~al.}(2013){Chaplin}, {Sanchis-Ojeda}, {Campante},
  {Handberg}, {Stello}, {Winn}, {Basu}, {Christensen-Dalsgaard}, {Davies},
  {Metcalfe}, {Buchhave}, {Fischer}, {Bedding}, {Cochran}, {Elsworth},
  {Gilliland}, {Hekker}, {Huber}, {Isaacson}, {Karoff}, {Kawaler}, {Kjeldsen},
  {Latham}, {Lund}, {Lundkvist}, {Marcy}, {Miglio}, {Barclay}, \&
  {Lissauer}}]{2013ApJ...766..101C}
{Chaplin}, W.~J., {et~al.} 2013, \apj, 766, 101

\bibitem[{{Collier Cameron}(2007)}]{2007AN....328.1030C}
{Collier Cameron}, A. 2007, Astronomische Nachrichten, 328, 1030

\bibitem[{{D{\'e}sert} {et~al.}(2011){D{\'e}sert}, {Charbonneau}, {Demory},
  {Ballard}, {Carter}, {Fortney}, {Cochran}, {Endl}, {Quinn}, {Isaacson},
  {Fressin}, {Buchhave}, {Latham}, {Knutson}, {Bryson}, {Torres}, {Rowe},
  {Batalha}, {Borucki}, {Brown}, {Caldwell}, {Christiansen}, {Deming},
  {Fabrycky}, {Ford}, {Gilliland}, {Gillon}, {Haas}, {Jenkins}, {Kinemuchi},
  {Koch}, {Lissauer}, {Lucas}, {Mullally}, {MacQueen}, {Marcy}, {Sasselov},
  {Seager}, {Still}, {Tenenbaum}, {Uddin}, \& {Winn}}]{2011ApJS..197...14D}
{D{\'e}sert}, J.-M., {et~al.} 2011, \apjs, 197, 14

\bibitem[{{Doyle} {et~al.}(1984){Doyle}, {Wilcox}, \&
  {Lorre}}]{1984ApJ...287..307D}
{Doyle}, L.~R., {Wilcox}, T.~J., \& {Lorre}, J.~J. 1984, \apj, 287, 307

\bibitem[{{Fabrycky} \& {Tremaine}(2007)}]{2007ApJ...669.1298F}
{Fabrycky}, D., \& {Tremaine}, S. 2007, \apj, 669, 1298

\bibitem[{{Gray}(2005)}]{2005oasp.book.....G}
{Gray}, D.~F. 2005, {The Observation and Analysis of Stellar Photospheres}, ed.
  {Gray, D.~F.}

\bibitem[{{H{\'e}brard} {et~al.}(2008){H{\'e}brard}, {Bouchy}, {Pont},
  {Loeillet}, {Rabus}, {Bonfils}, {Moutou}, {Boisse}, {Delfosse}, {Desort},
  {Eggenberger}, {Ehrenreich}, {Forveille}, {Lagrange}, {Lovis}, {Mayor},
  {Pepe}, {Perrier}, {Queloz}, {Santos}, {S{\'e}gransan}, {Udry}, \&
  {Vidal-Madjar}}]{2008A&A...488..763H}
{H{\'e}brard}, G., {et~al.} 2008, \aap, 488, 763

\bibitem[{{Hirano} {et~al.}(2011{\natexlab{a}}){Hirano}, {Narita}, {Sato},
  {Winn}, {Aoki}, {Tamura}, {Taruya}, \& {Suto}}]{2011PASJ...63L..57H}
{Hirano}, T., {Narita}, N., {Sato}, B., {Winn}, J.~N., {Aoki}, W., {Tamura},
  M., {Taruya}, A., \& {Suto}, Y. 2011{\natexlab{a}}, \pasj, 63, L57

\bibitem[{{Hirano} {et~al.}(2012{\natexlab{a}}){Hirano}, {Sanchis-Ojeda},
  {Takeda}, {Narita}, {Winn}, {Taruya}, \& {Suto}}]{2012ApJ...756...66H}
{Hirano}, T., {Sanchis-Ojeda}, R., {Takeda}, Y., {Narita}, N., {Winn}, J.~N.,
  {Taruya}, A., \& {Suto}, Y. 2012{\natexlab{a}}, \apj, 756, 66

\bibitem[{{Hirano} {et~al.}(2011{\natexlab{b}}){Hirano}, {Suto}, {Winn},
  {Taruya}, {Narita}, {Albrecht}, \& {Sato}}]{2011ApJ...742...69H}
{Hirano}, T., {Suto}, Y., {Winn}, J.~N., {Taruya}, A., {Narita}, N.,
  {Albrecht}, S., \& {Sato}, B. 2011{\natexlab{b}}, \apj, 742, 69

\bibitem[{{Hirano} {et~al.}(2012{\natexlab{b}}){Hirano}, {Narita}, {Sato},
  {Takahashi}, {Masuda}, {Takeda}, {Aoki}, {Tamura}, \&
  {Suto}}]{2012ApJ...759L..36H}
{Hirano}, T., {et~al.} 2012{\natexlab{b}}, \apjl, 759, L36

\bibitem[{{Hou} {et~al.}(2009){Hou}, {Parker}, {Harris}, \&
  {Wilman}}]{2009ApJ...702.1199H}
{Hou}, A., {Parker}, L.~C., {Harris}, W.~E., \& {Wilman}, D.~J. 2009, \apj,
  702, 1199

\bibitem[{{Huber} {et~al.}(2013){Huber}, {Carter}, {Barbieri}, {Miglio},
  {Deck}, {Fabrycky}, {Montet}, {Buchhave}, {Chaplin}, {Hekker},
  {Montalb{\'a}n}, {Sanchis-Ojeda}, {Basu}, {Bedding}, {Campante},
  {Christensen-Dalsgaard}, {Elsworth}, {Stello}, {Arentoft}, {Ford},
  {Gilliland}, {Handberg}, {Howard}, {Isaacson}, {Johnson}, {Karoff},
  {Kawaler}, {Kjeldsen}, {Latham}, {Lund}, {Lundkvist}, {Marcy}, {Metcalfe},
  {Silva Aguirre}, \& {Winn}}]{2013arXiv1310.4503H}
{Huber}, D., {et~al.} 2013, ArXiv e-prints

\bibitem[{{Kurucz}(1993)}]{1993KurCD..13.....K}
{Kurucz}, R. 1993, ATLAS9 Stellar Atmosphere Programs and 2 km/s grid.~Kurucz
  CD-ROM No.~13.~ Cambridge, Mass.: Smithsonian Astrophysical Observatory,
  1993., 13

\bibitem[{{Lai} {et~al.}(2011){Lai}, {Foucart}, \& {Lin}}]{2011MNRAS.412.2790L}
{Lai}, D., {Foucart}, F., \& {Lin}, D.~N.~C. 2011, \mnras, 412, 2790

\bibitem[{{Lin} {et~al.}(1996){Lin}, {Bodenheimer}, \&
  {Richardson}}]{1996Natur.380..606L}
{Lin}, D.~N.~C., {Bodenheimer}, P., \& {Richardson}, D.~C. 1996, \nat, 380, 606

\bibitem[{{Lissauer} {et~al.}(2012){Lissauer}, {Marcy}, {Rowe}, {Bryson},
  {Adams}, {Buchhave}, {Ciardi}, {Cochran}, {Fabrycky}, {Ford}, {Fressin},
  {Geary}, {Gilliland}, {Holman}, {Howell}, {Jenkins}, {Kinemuchi}, {Koch},
  {Morehead}, {Ragozzine}, {Seader}, {Tanenbaum}, {Torres}, \&
  {Twicken}}]{2012ApJ...750..112L}
{Lissauer}, J.~J., {et~al.} 2012, \apj, 750, 112

\bibitem[{{McQuillan} {et~al.}(2013{\natexlab{a}}){McQuillan}, {Aigrain}, \&
  {Mazeh}}]{2013MNRAS.432.1203M}
{McQuillan}, A., {Aigrain}, S., \& {Mazeh}, T. 2013{\natexlab{a}}, \mnras, 432,
  1203

\bibitem[{{McQuillan} {et~al.}(2013{\natexlab{b}}){McQuillan}, {Mazeh}, \&
  {Aigrain}}]{2013ApJ...775L..11M}
{McQuillan}, A., {Mazeh}, T., \& {Aigrain}, S. 2013{\natexlab{b}}, \apjl, 775,
  L11

\bibitem[{{Nagasawa} \& {Ida}(2011)}]{2011ApJ...742...72N}
{Nagasawa}, M., \& {Ida}, S. 2011, \apj, 742, 72

\bibitem[{{Naoz} {et~al.}(2011){Naoz}, {Farr}, {Lithwick}, {Rasio}, \&
  {Teyssandier}}]{2011Natur.473..187N}
{Naoz}, S., {Farr}, W.~M., {Lithwick}, Y., {Rasio}, F.~A., \& {Teyssandier}, J.
  2011, \nat, 473, 187

\bibitem[{{Narita} {et~al.}(2009){Narita}, {Sato}, {Hirano}, \&
  {Tamura}}]{2009PASJ...61L..35N}
{Narita}, N., {Sato}, B., {Hirano}, T., \& {Tamura}, M. 2009, \pasj, 61, L35

\bibitem[{{Narita} {et~al.}(2007){Narita}, {Enya}, {Sato}, {Ohta}, {Winn},
  {Suto}, {Taruya}, {Turner}, {Aoki}, {Yoshii}, {Yamada}, \&
  {Tamura}}]{2007PASJ...59..763N}
{Narita}, N., {et~al.} 2007, \pasj, 59, 763

\bibitem[{{Nutzman} {et~al.}(2011){Nutzman}, {Fabrycky}, \&
  {Fortney}}]{2011ApJ...740L..10N}
{Nutzman}, P.~A., {Fabrycky}, D.~C., \& {Fortney}, J.~J. 2011, \apjl, 740, L10

\bibitem[{{Ohta} {et~al.}(2005){Ohta}, {Taruya}, \&
  {Suto}}]{2005ApJ...622.1118O}
{Ohta}, Y., {Taruya}, A., \& {Suto}, Y. 2005, \apj, 622, 1118

\bibitem[{{Pont} {et~al.}(2010){Pont}, {Endl}, {Cochran}, {Barnes}, {Sneden},
  {MacQueen}, {Moutou}, {Aigrain}, {Alonso}, {Baglin}, {Bouchy}, {Deleuil},
  {Fridlund}, {H{\'e}brard}, {Hatzes}, {Mazeh}, \&
  {Shporer}}]{2010MNRAS.402L...1P}
{Pont}, F., {et~al.} 2010, \mnras, 402, L1

\bibitem[{{Press} \& {Rybicki}(1989)}]{1989ApJ...338..277P}
{Press}, W.~H., \& {Rybicki}, G.~B. 1989, \apj, 338, 277

\bibitem[{{Queloz} {et~al.}(2000){Queloz}, {Eggenberger}, {Mayor}, {Perrier},
  {Beuzit}, {Naef}, {Sivan}, \& {Udry}}]{2000A&A...359L..13Q}
{Queloz}, D., {Eggenberger}, A., {Mayor}, M., {Perrier}, C., {Beuzit}, J.~L.,
  {Naef}, D., {Sivan}, J.~P., \& {Udry}, S. 2000, \aap, 359, L13

\bibitem[{{Sanchis-Ojeda} {et~al.}(2011){Sanchis-Ojeda}, {Winn}, {Holman},
  {Carter}, {Osip}, \& {Fuentes}}]{2011ApJ...733..127S}
{Sanchis-Ojeda}, R., {Winn}, J.~N., {Holman}, M.~J., {Carter}, J.~A., {Osip},
  D.~J., \& {Fuentes}, C.~I. 2011, \apj, 733, 127

\bibitem[{{Sanchis-Ojeda} {et~al.}(2012){Sanchis-Ojeda}, {Fabrycky}, {Winn},
  {Barclay}, {Clarke}, {Ford}, {Fortney}, {Geary}, {Holman}, {Howard},
  {Jenkins}, {Koch}, {Lissauer}, {Marcy}, {Mullally}, {Ragozzine}, {Seader},
  {Still}, \& {Thompson}}]{2012Natur.487..449S}
{Sanchis-Ojeda}, R., {et~al.} 2012, \nat, 487, 449

\bibitem[{{Sanchis-Ojeda} {et~al.}(2013){Sanchis-Ojeda}, {Winn}, {Marcy},
  {Howard}, {Isaacson}, {Johnson}, {Torres}, {Albrecht}, {Campante}, {Chaplin},
  {Davies}, {Lund}, {Carter}, {Dawson}, {Buchhave}, {Everett}, {Fischer},
  {Geary}, {Gilliland}, {Horch}, {Howell}, \& {Latham}}]{2013ApJ...775...54S}
---. 2013, \apj, 775, 54

\bibitem[{{Schlaufman}(2010)}]{2010ApJ...719..602S}
{Schlaufman}, K.~C. 2010, \apj, 719, 602

\bibitem[{{Smith} {et~al.}(2012){Smith}, {Stumpe}, {Van Cleve}, {Jenkins},
  {Barclay}, {Fanelli}, {Girouard}, {Kolodziejczak}, {McCauliff}, {Morris}, \&
  {Twicken}}]{2012PASP..124.1000S}
{Smith}, J.~C., {et~al.} 2012, \pasp, 124, 1000

\bibitem[{{Steffen} {et~al.}(2012){Steffen}, {Ragozzine}, {Fabrycky}, {Carter},
  {Ford}, {Holman}, {Rowe}, {Welsh}, {Borucki}, {Boss}, {Ciardi}, \&
  {Quinn}}]{2012PNAS..109.7982S}
{Steffen}, J.~H., {et~al.} 2012, Proceedings of the National Academy of
  Science, 109, 7982

\bibitem[{{Stumpe} {et~al.}(2012){Stumpe}, {Smith}, {Van Cleve}, {Twicken},
  {Barclay}, {Fanelli}, {Girouard}, {Jenkins}, {Kolodziejczak}, {McCauliff}, \&
  {Morris}}]{2012PASP..124..985S}
{Stumpe}, M.~C., {et~al.} 2012, \pasp, 124, 985

\bibitem[{{Tajitsu} {et~al.}(2012){Tajitsu}, {Aoki}, \&
  {Yamamuro}}]{2012PASJ...64...77T}
{Tajitsu}, A., {Aoki}, W., \& {Yamamuro}, T. 2012, \pasj, 64, 77

\bibitem[{{Takeda} {et~al.}(2002){Takeda}, {Ohkubo}, \&
  {Sadakane}}]{2002PASJ...54..451T}
{Takeda}, Y., {Ohkubo}, M., \& {Sadakane}, K. 2002, \pasj, 54, 451

\bibitem[{{Takeda} {et~al.}(2005){Takeda}, {Ohkubo}, {Sato}, {Kambe}, \&
  {Sadakane}}]{2005PASJ...57...27T}
{Takeda}, Y., {Ohkubo}, M., {Sato}, B., {Kambe}, E., \& {Sadakane}, K. 2005,
  \pasj, 57, 27

\bibitem[{{Triaud} {et~al.}(2010){Triaud}, {Collier Cameron}, {Queloz},
  {Anderson}, {Gillon}, {Hebb}, {Hellier}, {Loeillet}, {Maxted}, {Mayor},
  {Pepe}, {Pollacco}, {S{\'e}gransan}, {Smalley}, {Udry}, {West}, \&
  {Wheatley}}]{2010A&A...524A..25T}
{Triaud}, A.~H.~M.~J., {et~al.} 2010, \aap, 524, A25+

\bibitem[{{Valenti} \& {Fischer}(2005)}]{2005ApJS..159..141V}
{Valenti}, J.~A., \& {Fischer}, D.~A. 2005, \apjs, 159, 141

\bibitem[{{Walkowicz} \& {Basri}(2013)}]{2013MNRAS.tmp.2426W}
{Walkowicz}, L.~M., \& {Basri}, G.~S. 2013, \mnras

\bibitem[{{Winn} {et~al.}(2010){Winn}, {Fabrycky}, {Albrecht}, \&
  {Johnson}}]{2010ApJ...718L.145W}
{Winn}, J.~N., {Fabrycky}, D., {Albrecht}, S., \& {Johnson}, J.~A. 2010, \apjl,
  718, L145

\bibitem[{{Winn} {et~al.}(2009){Winn}, {Johnson}, {Albrecht}, {Howard},
  {Marcy}, {Crossfield}, \& {Holman}}]{2009ApJ...703L..99W}
{Winn}, J.~N., {Johnson}, J.~A., {Albrecht}, S., {Howard}, A.~W., {Marcy},
  G.~W., {Crossfield}, I.~J., \& {Holman}, M.~J. 2009, \apjl, 703, L99

\bibitem[{{Winn} {et~al.}(2005){Winn}, {Noyes}, {Holman}, {Charbonneau},
  {Ohta}, {Taruya}, {Suto}, {Narita}, {Turner}, {Johnson}, {Marcy}, {Butler},
  \& {Vogt}}]{2005ApJ...631.1215W}
{Winn}, J.~N., {et~al.} 2005, \apj, 631, 1215

\bibitem[{{Wolf} {et~al.}(2007){Wolf}, {Laughlin}, {Henry}, {Fischer}, {Marcy},
  {Butler}, \& {Vogt}}]{2007ApJ...667..549W}
{Wolf}, A.~S., {Laughlin}, G., {Henry}, G.~W., {Fischer}, D.~A., {Marcy}, G.,
  {Butler}, P., \& {Vogt}, S. 2007, \apj, 667, 549

\bibitem[{{Wu} \& {Murray}(2003)}]{2003ApJ...589..605W}
{Wu}, Y., \& {Murray}, N. 2003, \apj, 589, 605

\bibitem[{{Yi} {et~al.}(2001){Yi}, {Demarque}, {Kim}, {Lee}, {Ree}, {Lejeune},
  \& {Barnes}}]{2001ApJS..136..417Y}
{Yi}, S., {Demarque}, P., {Kim}, Y.-C., {Lee}, Y.-W., {Ree}, C.~H., {Lejeune},
  T., \& {Barnes}, S. 2001, \apjs, 136, 417

\end{thebibliography}


\end{document}